\documentclass{article}

\usepackage{times}
\usepackage{graphicx} % more modern
\usepackage{subfigure} 

\usepackage{natbib}

\usepackage{algorithm}
\usepackage{algorithmic}

\usepackage{hyperref}

\usepackage[accepted]{icml2015}

\usepackage{amsbsy}

\newcommand{\T}{\mathrm{T}}
\newcommand{\E}{\mathrm{E}}

\newcommand{\Tr}{\mathrm{Tr}}

\newcommand{\norm}{\mathcal{N}}

\newcommand{\bvect}{\mathbf{b}}
\newcommand{\dvect}{\mathbf{d}}
\newcommand{\evect}{\mathbf{e}}
\newcommand{\rvect}{\mathbf{r}}

\newcommand{\svect}{\mathbf{s}}

\newcommand{\zvect}{\mathbf{z}}
\newcommand{\xvect}{\mathbf{x}}
\newcommand{\yvect}{\mathbf{y}}

\newcommand{\thetavect}{\boldsymbol{\theta}}
\newcommand{\deltavect}{\boldsymbol{\delta}}

\newcommand{\etavect}{\boldsymbol{\eta}}

\icmltitlerunning{Enabling scalable stochastic gradient-based inference for Gaussian processes}

\begin{document} 

\twocolumn[
\icmltitle{Enabling scalable stochastic gradient-based inference for Gaussian processes by employing the Unbiased LInear System SolvEr (ULISSE)}

%% Solving Linear Systems Unbiasedly to Enable Scalable Stochastic Gradient based Inference for Gaussian Processes

%% Solving Linear Systems Unbiasedly and 
%% Estimating Unbiased Solution of Linear Systems

%% Scaling GPs without deterministic approximations
%% No Deterministic Approximations - NO DA 
%% Accelerating Bayesian Computations - ABACO
%% Stochastic Gradient - S-G
%% Stochastic Gradient based Inference - SGBI - ESTO-GRABI
%% Gradient Based - GRAB
%% Unbiased Solution of Linear Systems - USOLS - UNSOLS - USO-LIS
%% Solving Linear Systems Unbiasedly - SOLSU - SOLLINSU
%% Enabling - E
%% Bayesian Inference - BI
%% Fast - F
%% Early Stop Unbiased Estimation - ESUES - ESTUNE
%% Scaling Bayesian Computations - SBC
%% Gaussian Processes - GAP - GP
%% Inference of Covariance Parameters - IOCP - INOCOPA

% unless you've provided the [accepted] option to the icml2015
\icmlauthor{Maurizio Filippone}{Maurizio.Filippone@glasgow.ac.uk}
\icmladdress{School of Computing Science,
            University of Glasgow, UK}
\icmlauthor{Raphael Engler}{Raphael.Engler@web.de}
\icmladdress{School of Computing Science,
            University of Glasgow, UK}

\icmlkeywords{Gaussian Processes, Stochastic Gradients, Bayesian Inference, Scalable Inference}

\vskip 0.3in
]

\begin{abstract} 
In applications of Gaussian processes where quantification of uncertainty is of primary interest, it is necessary to accurately characterize the posterior distribution over covariance parameters.
This paper proposes an adaptation of the Stochastic Gradient Langevin Dynamics algorithm to draw samples from the posterior distribution over covariance parameters with negligible bias and without the need to compute the marginal likelihood.
In Gaussian process regression, this has the enormous advantage that stochastic gradients can be computed by solving linear systems only.
A novel unbiased linear systems solver based on parallelizable covariance matrix-vector products is developed to accelerate the unbiased estimation of gradients.
The results demonstrate the possibility to enable scalable and exact (in a Monte Carlo sense) quantification of uncertainty in Gaussian processes without imposing any special structure on the covariance or reducing the number of input vectors.
\end{abstract} 

\section{Introduction}
\label{sec:introduction}

Probabilistic kernel machines based on Gaussian Processes (GPs) \cite{Rasmussen06} are popular in a number of applied domains as they offer the possibility to flexibly model complex data and, depending on the choice of covariance function, to gain some understanding on the underlying behavior of the system under study. 
When quantification of uncertainty is of primary interest, it is necessary to accurately characterize the posterior distribution over covariance parameters.
This has been argued in a number of papers where this is done by means of Markov chain Monte Carlo (MCMC) methods \cite{Williams95,Williams98,Neal99,Murray10,Taylor12,FilipponeML13,FilipponeIEEETPAMI14}.
%% A number of contributions have been proposed on how to carry out inference over GP covariance parameters by means of Markov chain Monte Carlo (MCMC) methods \cite{Williams98}, \cite{Murray10}, \cite{FilipponeML13}, \cite{FilipponeIEEETPAMI14}.

The limitation of MCMC approaches to draw samples from the posterior distribution over covariance parameters is that they need to compute the marginal likelihood at every iteration.
In GP regression, a standard way to compute the marginal likelihood involves storing and factorizing an $n \times n$ matrix, leading to $O(n^3)$ time and $O(n^2)$ space complexities, where $n$ is the size of the data set.
For large data sets this becomes unfeasible, so a large number of contributions can be found in the literature on how to make these calculations tractable. % by exploiting special structures of the covariance or reducing the number of input vectors.
For example, when the GP covariance matrix has some particular properties, e.g., it has sparse inverse \cite{Rue09,Simpson13,Lyne14}, it is computed on regularly spaced inputs \cite{Saatci11}, or it is computed on univariate inputs \cite{Gilboa15}, it is possible to considerably reduce the complexity in computing the marginal likelihood. % thus yielding the possibility to scale inference to millions of input vectors.
When these properties do not hold, which is common in several Machine Learning applications, approximations are usually employed.
Some examples involve the use subsets of the data \cite{Candela05}, the determination of a small number of surrogate input vectors that represent the full set of inputs \cite{Titsias09,Hensman13}, and the application of GPs to subsets of the data obtained by partitioning the input space \cite{Gramacy04}, to name a few. 
Unfortunately, it is difficult to assess to what extent approximations affect the quantification of uncertainty in predictions, although some interesting results in this direction are reported in \cite{Banerjee12}.

The focus of this paper are applications of GP regression where the structure of the covariance matrix is not necessarily special and quantification of uncertainty is of primary interest, so that approximations should be avoided.
%% Our first attempt to scale computations for GPs followed a recent trend in statistics where computations are approximated in an unbiased fashion and used in a way that ensure exactness, in a Monte Carlo sense, of the inference results.
%% Approaches along these lines have been proposed in inference for GPs with non-Gaussian likelihoods \cite{FilipponeIEEETPAMI14}, \cite{FilipponeICPR14}, sampling from high-dimensional Gaussians \cite{Simpson13}.
%% In the case of GP regression this would amount in obtaining an unbiased estimate of the marginal likelihood which involves estimating the log-determinant of the covariance matrix.
%% These calculations can be done using iterative methods involving covariance matrix-vector products \cite{Chen11}, \cite{Antinescu12}, \cite{Higham08}, \cite{Skilling93}, \cite{Seeger00}.
%% However, this is extremely computational demanding, as the convergence of these methods is slow and each iterations requires several covariance matrix-vector products.
%% Also, it is necessary to transform the unbiased version of the log-likelihhod into an unbiased version of the likelihood which introduces a number of other issues \cite{Girolami14}.
This paper proposes an adaptation of the Stochastic Gradient Langevin Dynamics (SGLD) algorithm \cite{Welling11} to draw samples from the posterior distribution over GP covariance parameters.
SGLD does not require the computation of the marginal likelihood and yields samples from the posterior distribution of interest with negligible bias. 
This has the enormous advantage that stochastic gradients can be computed by solving linear systems only \cite{GibbsPhD97,Gibbs97,Stein13}.
%% Therefore, we decided to employ the stochastic gradient based inference method in \cite{Welling11} that does not require the computation of the marginal likelihood but only of a noisy version of its gradient.
Linear systems can be solved by means of iterative methods, such as the Conjugate Gradient (CG) algorithm, that are based on parallelizable covariance matrix-vector products \cite{Higham08,Skilling93,Seeger00}.
Similar ideas were previously put forward to optimize GP covariance parameters \cite{Chen11,Antinescu12,Stein13}.
Despite the $O(n^2)$ in time and $O(n)$ in space complexities of these methods compare well with the $O(n^3)$ in time and $O(n^2)$ in space complexities of traditional MCMC-based inference, solving dense linear systems at each iteration makes the whole inference framework too slow to be of practical use.
We compare a number of standard ways to speed up the solution of dense linear systems, such as fast covariance matrix-vector products \cite{Gray00,Moore00} and preconditioning \cite{Srinivasan14}, and in line with what reported in \cite{Murray09}, we observe that they yield little gain in computational speed compared to the standard CG algorithm.
In order to enable practical inference for GPs applied to large data sets, we therefore develop an Unbiased LInear Systems SolvEr (ULISSE) that essentially allows the CG algorithm to stop early while retaining unbiasedness of the solution.
%% Motivated by the necessity to enable practical inference for GPs applied to large data sets, we develop an Unbiased LInear Systems SolvEr (ULISSE) that essentially allows the CG algorithm to stop early while retaining unbiasedness of the solution.

We highlight here that (i) in \cite{Welling11}, an unbiased estimate of the gradient is computed by considering small batches of data.
Recent alternative contributions on scaling Bayesian inference by analyzing small batches of data can be found in \cite{Banterle14,Maclaurin14}. 
GPs do not lend themselves to this treatment, due to the covariance structure making all data dependent on one another.
(ii) ULISSE is complementary to recent approaches in the area of probabilistic numerics that aim at infering, rather than computing, solutions to linear systems \cite{Hennig14}.
(iii) The proposed inference method is based on ``noisy'' gradients and is complementary to recent inference approaches based on noisy likelihoods \cite{Lyne14,FilipponeICPR14}.
In GP regression, iterative methods akin to the CG algorithm \cite{Higham08} can be employed to obtain an unbiased estimate of the log-determinant of the covariance matrix, but this remains an extremely onerous calculation needed to get an unbiased estimate of the log-marginal likelihood.
A further and perhaps more challenging issue is transforming the unbiased estimate of the log-marginal likelihood in an unbiased estimate of the marginal likelihood %, as this introduces further variance in the estimate and potentially leads to negative estimates of the marginal likelihood 
\cite{Kennedy85,Liu00,Lyne14}.
%% In this paper we take this observation one step forward and we study ways to speed up the solution of linear systems in order to make the sampling of covariance parameters from their posterior distribution practical for large datasets.

This paper demonstrates that employing ULISSE within SGLD makes it possible to accurately carry out inference of covariance parameters in GPs and effectively scale these computations to large data sets.
We report results on a data set with about $23$ thousand input vectors where we can draw ten thousand samples per day from the posterior distribution over covariance parameters on a desktop machine with standard hardware\footnote{Code to reproduce all the results can be found here: \\\url{www.dcs.gla.ac.uk/~maurizio/pages/code.html}}. 
To the best of our knowledge, this paper reports the first real attempt to enable full quantification of uncertainty of covariance parameters of GPs without reducing the number of input vectors and without imposing sparsity on the GP covariance or its inverse.

The paper is organized as follows.
Section~\ref{sec:inference} briefly reviews GPs and motivates the adoption of SGLD to infer GP covariance parameters.
Section~\ref{sec:linear:system} describes and evaluates the CG algorithm to solve linear systems and some variants based on fast covariance matrix-vector product and preconditioning.
Section~\ref{sec:ulisse} presents ULISSE and its use to obtain an unbiased estimate of the gradient of the log-marginal likelihood in GPs.
Section~\ref{sec:experiments} demonstrates the ability of the proposed methodology to accurately infer covariance parameters in GPs and its scalability properties to a large data set where the marginal likelihood cannot be computed exactly.
Finally, Section~\ref{sec:conclusions} draws the conclusions.

\section{Inference of covariance parameters in GPs} \label{sec:inference}

In GP regression, a set of continuous labels $\yvect = \{y_1, \ldots, y_n\}$ is associated with a set of input vectors $X = \{ \xvect_1, \ldots, \xvect_n \}$.
%% , and assume
%% \begin{equation}
%% y_i = f(\xvect_i) + \epsilon_i
%% \end{equation}
%% where the noise $\epsilon_i$ is i.i.d. and Gaussian distributed $\norm(\varepsilon_i | 0, \lambda)$.
%% One way to avoid specifying a parametric form for the unobserved function $f(\cdot)$ is to assume a nonparametric specification using a GP prior.
%% The GP prior is a prior over functions and it is formally defined as a set of random variables such that any of its subsets are jointly Gaussian distributed.
%% These variables are conveniently indexed by the elements of the domain of the input space.
%% GPs are characterized by the way the mean of these random variables and their covariance structure are specified.
%% Without loss of generality, in this work we will employ zero mean GPs.
%% In regression problems, GPs are employed to model a set of latent (unobserved) variables associated with each of the input vectors.
%% Under the assumption that continuous labels can be modeled by means of a Gaussian distribution centered at the latent variables, it is possible to integrate the latent variables out of the model and model directly the distribution over the labels $\yvect$ as another GP. 
%% For example, in the case of the Radial Basis Function (RBF) covariance function, we obtain
Throughout the paper, we will employ zero mean GPs with the following covariance function:
\begin{equation} \label{eq:covariance:rbf}
k(\xvect_i, \xvect_j) = \sigma \exp\left( \tau \| \xvect_i - \xvect_j  \|^2  \right) + \lambda \delta_{ij}
\end{equation}
%% Intuitively the RBF function assigns high covariance to pairs of variables associated with inputs that are close to each other in the input space and low covariance to variables associated with inputs that are far apart.
with $\delta_{ij} = 1$ if $i=j$ and zero otherwise. 
The parameter $\tau$ determines the rate of decay of the covariance function, whereas $\sigma$ represents the marginal variance of each Gaussian random variable comprising the GP.
The parameter $\lambda$ is the variance of the (Gaussian) noise on the labels.
Let $K$ be the covariance matrix with $K_{ij} = k(\xvect_i, \xvect_j)$
%% In order to apply GPs to the regression problem above, we observe that 
%% The marginal distribution of the GP modeling the unobserved function at the inputs spcified in $X$ will be multivariate Gaussian with zero mean and covariance matrix $Q$, where each element $Q_{ij} = q(\xvect_i, \xvect_j)$.
%% A further observation is that the Gaussian assumption on $\varepsilon_i$ coupled with the GP prior on the latent function allows one to integrate out the latent function and model the joint distribution over $\yvect$ directly as a GP with zero mean and covariance given by $K = Q + \lambda I$.
and denote by $\thetavect$ the vector comprising all parameters of the covariance matrix $K$, namely $\thetavect = (\sigma, \tau, \lambda)$.
%% Such parameters need to be optimized or inferred in order to fit the data and to be able to accurately predict the distribution over the labels associated with new input vectors.
%% One way to optimize the parameters would be to carry out the maximization of the so called marginal likelihood $p(\yvect | \thetavect, X)$ (possibly multiplied by the prior). % wrt $\thetavect$.
%% %% These would yield the Maximum Likelihood or the Maximum A Posteriori (MAP) solutions.
%% However, optimization does not take into account any uncertainty in the estimates of the parameters, and this could potentially affect the ability of the model to quantify uncertainty in predictions.
%% %% This has been widely discussed for GPs in \cite{Neal99,Rue09,FilipponeIEEETPAMI14,Taylor12}.

In a Bayesian sense, we would like to carry any uncertainty in parameters estimates forward to predictions over the label $y_*$ for a new input vector $\xvect_*$.
%% In particular, this would amount in integrating out covariance parameters from the model in order to compute the predictive distribution:
In particular, this requires solving the following integral:
\begin{equation}
p(y_* | \yvect, X, \xvect_*) = \int p(y_* | \yvect, \thetavect, X, \xvect_*) p(\thetavect | \yvect, X) d\thetavect.
%% p(y_* | \yvect, X, \xvect_*) = \int p(y_* | \yvect, X, \xvect_*, \thetavect) p(\thetavect | \yvect, X) d\thetavect.
\end{equation}
%% where we have dropped the conditioning on $X$ and $\xvect_*$ to keep the notation uncluttered.
%% This shows that in order to compute the predictive distribution for new input vectors, we need to characterize the posterior distribution over the covariance parameters $p(\thetavect | \yvect, X)$.
%% The integral above is analytically intractable, so it is necessary to resort to some approximations.
Such an expectation, like any other expectation under the posterior over $\thetavect$, is analytically intractable, so it is necessary to resort to some approximations.
A standard way to tackle this intractability is to draw samples from $p(\thetavect | \yvect, X)$ using MCMC methods, and approximate the expectation with the Monte Carlo estimate
\begin{equation} \label{eq:monte:carlo:integration}
p(y_* | \yvect, X, \xvect_*) \simeq \frac{1}{N} \sum_{i=1}^N p(y_* | \thetavect^{(i)}, X, \xvect_*) ,
\end{equation}
where $\thetavect^{(i)}$ denotes the $i$th of a set of samples from $p(\thetavect | \yvect, X)$.
%% This estimate will asymptotically converge to the exact expectation $p(y_* | \yvect)$.
Drawing samples from the posterior distribution can be done using several MCMC algorithms that essentially are based on a proposal mechanism and on an accept/reject step that requires the evaluation of the log-marginal likelihood: % $p(\yvect | \thetavect, X)$.
\begin{equation}
\log[p(\yvect | \thetavect, X)] = -\frac{1}{2} \log\left(|K|\right) - \frac{1}{2} \yvect^{\T} K^{-1} \yvect + \mathrm{const.}
\end{equation}
A standard way to proceed, is to factorize the covariance matrix $K = L L^{\T}$ using the Cholesky algorithm \cite{Golub96}.
The factorization costs $O(n^3)$ operations and requires the storage of $O(n^2)$ entries of the covariance matrix, but after that computing the log-determinant and the inverse of $K$ multiplied by $\yvect$ can be done using $O(n^2)$ operations.

The computational complexities above pose a constraint on the scalability of GPs to large data sets.
Iterative methods based on covariance matrix-vector products (CMVPs) have been proposed to obtain an unbiased estimate of the log-marginal likelihood. %log-determinant term.
%% In the literature, alternative ways to unbiasedly estimate the log-marginal likelihood have been proposed, and are based on iterative methods based on covariance matrix-vector products (CMVPs) \cite{Kennedy85,Liu00,Chen11,Antinescu12,Stein13}.
%% Even though these methods scale in $O(n^2)$ in time and $O(n)$ in space the number of iterations to reach convergence in estimating the log-determinant term can be prohibitively large.
Even though these methods scale with $O(n^2)$ in time and $O(n)$ in space, they are of little practical use in the task of sampling from $p(\thetavect | \yvect, X)$, as the number of iterations needed to estimate the log-determinant term can be prohibitively large (see, e.g., \cite{Chen11}). % and this represents an obstacle preventing these techniques to be of practical use in the sampling from the posterior distribution of covariance parameters.
%% A further complication is how to make effective use of an unbiased estimate of the log-marginal likelihood in noisy Monte Carlo algorithm \cite{Lyne14}. 
We now illustrate our proposal to obtain samples from $p(\thetavect| \yvect, X)$ with negligible bias and without having to estimate log-determinants and marginal likelihoods.

\subsection{Stochastic Gradient Langevin Dynamics (SGLD)}

%% In order to overcome the challenges associated with the estimation of $\log|K|$ while retaining a fully Bayesian inference engine without deterministic approximations, 
We briefly describe how to adapt SGLD \cite{Welling11} to obtain samples from $p(\thetavect | \yvect, X)$ in GPs.
The idea behind SGLD is to modify the standard stochastic gradient optimization algorithm \cite{Robbins51} by injecting Gaussian noise in a way that ensures transition into a Langevin dynamics phase yielding samples from the posterior distribution of interest.
In particular, the proposal of a new set of parameters is
\begin{equation}
\thetavect_{t+1} = \thetavect_{t} + \frac{\varepsilon_t}{2} M \left\{ \tilde{\mathbf{g}} + \nabla_{\thetavect} \log[p(\thetavect)] \right\} + \etavect_t
\end{equation}
with $\etavect_t \sim \norm(\etavect_t | 0, \varepsilon_t M)$ 
and $\tilde{\mathbf{g}}$ an unbiased estimate of the gradient of $\log[p(\yvect | \thetavect, X)]$.
%% The term in curly brackets is an unbiased estimate of $\log[p(\thetavect | \yvect, X)]$.
We have also introduced a preconditioning matrix $M$ that can be chosen to improve convergence of SGLD.
The update equation, except for $\etavect_t$, is the standard update used in stochastic gradient optimization.
The parameters $\varepsilon_t$ are chosen to satisfy
\begin{equation}
\sum_{t=1}^{\infty} \varepsilon_t = \infty
\quad
\mathrm{and}
\quad
\sum_{t=1}^{\infty} \varepsilon_t^2 < \infty
\end{equation}
as these conditions, along with some other technical assumptions, guarantee convergence to a local maximum.
%% In this work, we define $\varepsilon_t = a(b + t)^{-\gamma}$ with $\gamma = 1$. %$\in (0.5, 1]$.
The injected noise $\etavect_t$ is Gaussian with covariance $\varepsilon_t M$ ensuring that the algorithm transitions into a discretized version of a Langevin dynamics with target distribution given by the posterior over $\thetavect$.
In principle, it would be necessary to accept or reject the proposals, % made during the Langevin dynamics phase, 
which would require evaluating the marginal likelihood.
The key result in \cite{Welling11} is that when SGLD reaches the Langevin dynamics phase, the step-size $\varepsilon_t$ is small enough to make the acceptance rate close to one.
Therefore, in this phase it is possible to accept all proposals, avoiding having to evaluate the marginal likelihood, at the cost of introducing a negligible amount of bias.
%% Interestingly, as the step-size decreases the acceptance rate asymptotically tends to one and the effective sample size tends to infinity.

%% We can estimate when the algorithm reaches the Langevin dynamics phase.
Following \cite{Welling11}, we can estimate when the algorithm reaches the Langevin dynamics phase by monitoring the ratio between the variance of the stochastic gradients and the variance of the injected noise.
Defining $V$ to be the sampling covariance of the stochastic gradients and $\lambda_{\max}(A)$ to be the largest eigenvalue of a matrix $A$, we can write such a ratio as
\begin{equation}
\frac{\varepsilon_t}{4} \lambda_{\max}\left( M^{\frac{1}{2}} V M^{\frac{1}{2}}\right)
\end{equation}
When this ratio is small enough the algorithm is in its Langevin dynamics phase and produces samples from the posterior distribution over $\thetavect$.
Further theoretical analyses on the convergence properties of SGLD can be found in \cite{Teh14,Vollmer15}.

The motivation for employing SGLD for inference of GP covariance parameters comes from inspecting the gradient of the log-marginal likelihood that has components % $g_i = \frac{\partial \log[p(\yvect | \thetavect, X)]}{\partial \theta_i}$:
\begin{equation} \label{eq:gradient:exact}
%% \frac{\partial \log[p(\yvect | \thetavect)]}{\partial \thetavect_i} = 
%% - \frac{1}{2} \Tr \left( K^{-1} \frac{\partial K}{\partial \thetavect_i} \right) + \yvect^{\T} K^{-1} \frac{\partial K}{\partial \thetavect_i} K^{-1} \yvect
%% \frac{\partial \log[p(\yvect | \thetavect)]}{\partial \theta_i} = 
g_i = 
- \frac{1}{2} \Tr \left( K^{-1} \frac{\partial K}{\partial \theta_i} \right) + \frac{1}{2} \yvect^{\T} K^{-1} \frac{\partial K}{\partial \theta_i} K^{-1} \yvect
\end{equation}
Computing the $g_i$'s requires again $O(n^3)$ operations due to the trace term and the linear system $K^{-1} \yvect$.
However, we can introduce $N_{\rvect}$ vectors $\rvect^{(i)}$ with components drawn from $\{-1, 1\}$ with probability $1/2$ and unbiasedly estimate the trace term \cite{GibbsPhD97}, obtaining:
\begin{equation} \label{eq:gradient:unbiased}
\tilde{g}_i = 
%% - \frac{1}{2} \rvect^{\T} K^{-1} \frac{\partial K}{\partial \theta_i} \rvect + \yvect^{\T} K^{-1} \frac{\partial K}{\partial \theta_i} K^{-1} \yvect
- \frac{1}{2 N_{\rvect}} \sum_{i=1}^{N_{\rvect}} {\rvect^{(i)}}^{\T} K^{-1} \frac{\partial K}{\partial \theta_i} \rvect^{(i)} + \frac{1}{2}\yvect^{\T} K^{-1} \frac{\partial K}{\partial \theta_i} K^{-1} \yvect
\end{equation}
Given that $\E( \rvect^{(i)} {\rvect^{(i)}}^{\T} ) = I$,
we can readily verify that % the expectation of each of the terms in the sum is 
$
\E[{\rvect^{(i)}}^{\T} K^{-1} \frac{\partial K}{\partial \theta_i} \rvect^{(i)}] = 
%% \E[K^{-1} \frac{\partial K}{\partial \theta_i} \rvect^{(i)} {\rvect^{(i)}}^{\T} ] =
\Tr\left[K^{-1} \frac{\partial K}{\partial \theta_i} \E( \rvect^{(i)} {\rvect^{(i)}}^{\T} ) \right]
$, 
which yields the trace term in eq.~\ref{eq:gradient:exact}.
Hence, in order to compute an unbiased version of the gradient of the log-marginal likelihood we need to solve one linear system for $\yvect$ and one for each of the $N_{\rvect}$ vectors $\rvect^{(i)}$ used to estimate the trace term.
This consideration forms the basis of the proposed methodology.
Computing an unbiased version of the gradient involves solving linear systems only, which is much easier and cheaper than estimating log-determinants.
%% We now discuss how to solve linear systems in $O(n^2)$ time and $O(n)$ space. %, and in section~\ref{sec:ulisse} we develop a fast unbiased solver of linear systems that builds .
%% The focus of this paper is on how to scale the computations associated with the solution of dense linear systems and how these ideas can practially contribute to scale inference in GPs.

\section{Solving linear systems without storing $K$} \label{sec:linear:system}

We have discussed that SGLD to infer covariance parameters in GPs requires solving linear systems.
%% The key to make this methodology practical is therefore being able to solve linear systems quickly.
Here we briefly review the Conjugate Gradient (CG) algorithm that is a popular method to iteratively solve linear systems based on Covariance Matrix Vector Product (CMVP) operations.
CMVPs can be carried out without having to store $K$, so their time and space complexities are in $O(n^2)$ and $O(n)$, respectively.
We also discuss and evaluate a few variants to speed up computations/convergence, such as preconditioning and fast CMVPs.
Throughout this section we will evaluate the effectiveness of these alternatives on a GP regression task applied to the Concrete data set from the UCI repository \cite{Asuncion07}.
This data set contains data about the compressive strength of $n=1030$ samples of concrete described by $d = 8$ features. % such as age and weight of ingredients. % (input matrix $X$).
%% The dataset contains $n=1030$ input vectors.
%% The analysis of these alternatives indicates that there is little (if any) gain in using preconditioning and fast CMVPs and motivates the need to develop an alternative way to solve dense linear systems, as presented in Section~4.

\subsection{The Conjugate Gradient (CG) algorithm}

\begin{algorithm}[tb]
   \caption{The Conjugate Gradient algorithm}
   \label{alg:CG:algorithm}
\begin{algorithmic}
   \STATE {\bfseries Input:} data $X$, vector $\bvect$, convergence threshold $\epsilon$, initial vector $\svect_0$, maximum number of iterations $T$
   \STATE $\evect_0 = \bvect - K \svect_0$; \quad $\dvect_0 = \evect_0$;
   \FOR{$i=0$ {\bfseries to} $T$}
   \STATE $\displaystyle \alpha_i = \frac{\evect_i^{\T} \evect_i}{\dvect_i^{\T} K \dvect_i}$;
   \STATE $\svect_{i+1} = \svect_i + \alpha_i \dvect_i$;
   \STATE $\evect_{i+1} = \evect_i - \alpha_i K \dvect_i$;
   \IF{$\| \evect_{i+1} \| < \epsilon$}
   \STATE{return $\svect = \svect_{i+1}$;}
   \ENDIF
   \STATE $\displaystyle \beta_i = \frac{\evect_{i+1}^{\T} \evect_{i+1}}{\evect_i^{\T} \evect_i}$;
   \STATE $\dvect_{i+1} = \evect_{i+1} + \beta_i \dvect_i$;
   \ENDFOR
\end{algorithmic}
\end{algorithm}

Given a linear system of the form $K \svect = \bvect$ with $K$ and $\bvect$ given, the CG algorithm \cite{Golub96} yields the solution $\svect$ without having to invert or factorize the matrix $K$.
The idea is to calculate the solution $\svect$ as the minimizer of
\begin{equation}
\phi(\svect) = \frac{1}{2} \svect^{\T} K \svect - \svect^{\T} \bvect
\end{equation}
which can be obtained by employing gradient-based optimization.
%% The CG algorithm is characterized by the fact that a sequence of orthornormal vectors is generated iteratively and at each iteration the solution $\svect$ is updated by minimizing $\phi$ over the set spanned by the orthonormal vectors.
%% The CG algorithm is derived so that the calculation of the orthonormal vectors and the update of $\svect$ do not require expensive operations.
The CG algorithm is initialized from an initial guess $\svect_0$.
After that, the iterations refine the solution $\svect$ by updates in directions $\dvect_i$.
The CG algorithm, in comparison with the standard gradient descent, is characterized by the fact that $K$-orthogonality (or conjugacy with respect to $K$) of the search directions is imposed, namely $\dvect_i^{\T} K \dvect_j = 0$ when $i \neq j$. 
This condition yields a sequence of residuals $\evect_i = \bvect - K \svect_i$ that are mutually orthogonal, and guarantees convergence in at most $n$ iterations. 
Remarkably, the CG algorithm can be implemented in a way that requires a single CMVP ($K \dvect_i$) at each iteration (see Algorithm~\ref{alg:CG:algorithm}).
%% The CG algorithm is sketched in Algorithm~\ref{alg:CG:algorithm}.

\begin{figure}[t]
% \vskip 0.2in
\begin{center}
\centerline{\includegraphics[width=\columnwidth]{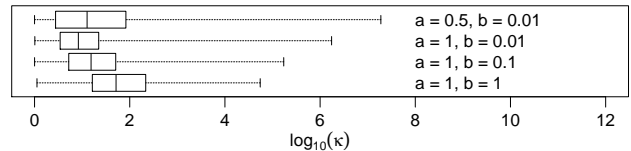} }
\caption{Distribution of the condition number $\kappa$ of the covariance matrix for different choices of shape and rate parameters of a Gamma prior on each covariance parameter $\theta$.}
\label{fig:condition:number:prior}
\end{center}
\vskip -0.2in
\end{figure}

The trade-off between accuracy and speed is governed by the threshold $\epsilon$, which in this paper is set to $\epsilon = 10^{-8}$.
%% The threshold $\epsilon$ can be chosen to tradeoff accuracy for speed; in this paper we set $\epsilon = 10^{-8}$.
Theoretically, the CG algorithm is guaranteed to converge in at most $n$ iterations, but in practice, due to the representation in finite numerical precision, orthogonality of the directions can be lost, especially in badly conditioned systems, and the CG algorithm can take more than $n$ iterations to converge.
%% This consideration suggests that the CG algorithm should be implemented using high precision operations in order to preserve orthogonality between the directions $\dvect_i$, and that employing fast CMVPs might severely affect convergence. % and that would lead to faster convergence.
%% %% This suggests that it is necessary to find a tradeoff between computational cost of carrying out CMVPs and converge speed of the CG algorithm.
%% %% We aim to investigate to what extent employing fast CMVPs affects convergence.
%% Another way to improve convergence speed of the CG algorithm is based on the use of preconditioning techniques. 
%% We aim to investigate the use of preconditioners in the solution of linear systems involving the kernel matrix keeping the constraint that we can only afford CMVPs.
%% The first analysis that we report here, is the condition number of the covariance matrix of a GP modeling the mapping between inputs and labels.
The condition number of a matrix is defined as the ratio between its largest and smallest eigenvalues:
$$
\kappa = \frac{\lambda_{\max}(K)}{\lambda_{\min}(K)}
$$
Fig.~\ref{fig:condition:number:prior} shows the distribution of the condition number when each covariance parameter $\theta_i$ is sampled form a Gamma distribution with shape and rate parameters $a$ and $b$.
The distributions are reasonably vague and give a rough idea of the typical condition numbers encountered during the inference of GP covariance parameters for the Concrete data set.
%% We can expect that in most cases the posterior distribution over the covariance parameters will be more concentrated than the prior, especially if the priors are quite vague as the ones reported in the figure.
%% Therefore we can reasonably assume that the condition number of the covariances encountered during the sampling process will not exceed the ones for the given prior, although some proposals might end up in regions of the space what are not covered by the prior.
We can expect slower convergence speed when the condition number is large due to numerical instabilities; we are interested in quantifying to what extent this applies to GPs and what is the impact of cheap CMVPs and preconditioning on convergence speed.
%% Before briefly describing fast multipole methods and preconditioning for linear systems we emphasize that we are interested in assessing the impact of these alternative ways of solving linear systems on convergence speed of the CG algorithm and precision of the solution.
In the remainder of this section, we will consider the problem of solving the linear system $K \svect = \yvect$ that is needed in the calculation of part of the gradient in eq.~\ref{eq:gradient:unbiased}. 
The results pertaining to the solution of the linear systems $K \svect = \rvect^{(i)}$ are quite similar, so for the sake of brevity we will omit them. 

\subsection{Fast CMVPs}

%% The expensive operation in each iteration of the CG algorithm is the CMVP one.
%% Recently, there have been a number of contributions in the literature that aim at speeding up these operations at the price of a lower accuracy.
We consider here the use of two fast CMVPs based on efficient representation of input data that we will call ``kdtree'' \cite{Gray00} and ``anchors'' \cite{Moore00}\footnote{code implementing these methods can be found here: \url{www.cs.ubc.ca/~awll/nbody_methods.html}}. 
These methods yield fast CMVPs at the price of a lower accuracy.

\begin{figure}[t]
%% \vskip 0.2in
\begin{center}
\centerline{\includegraphics[width=0.49\columnwidth]{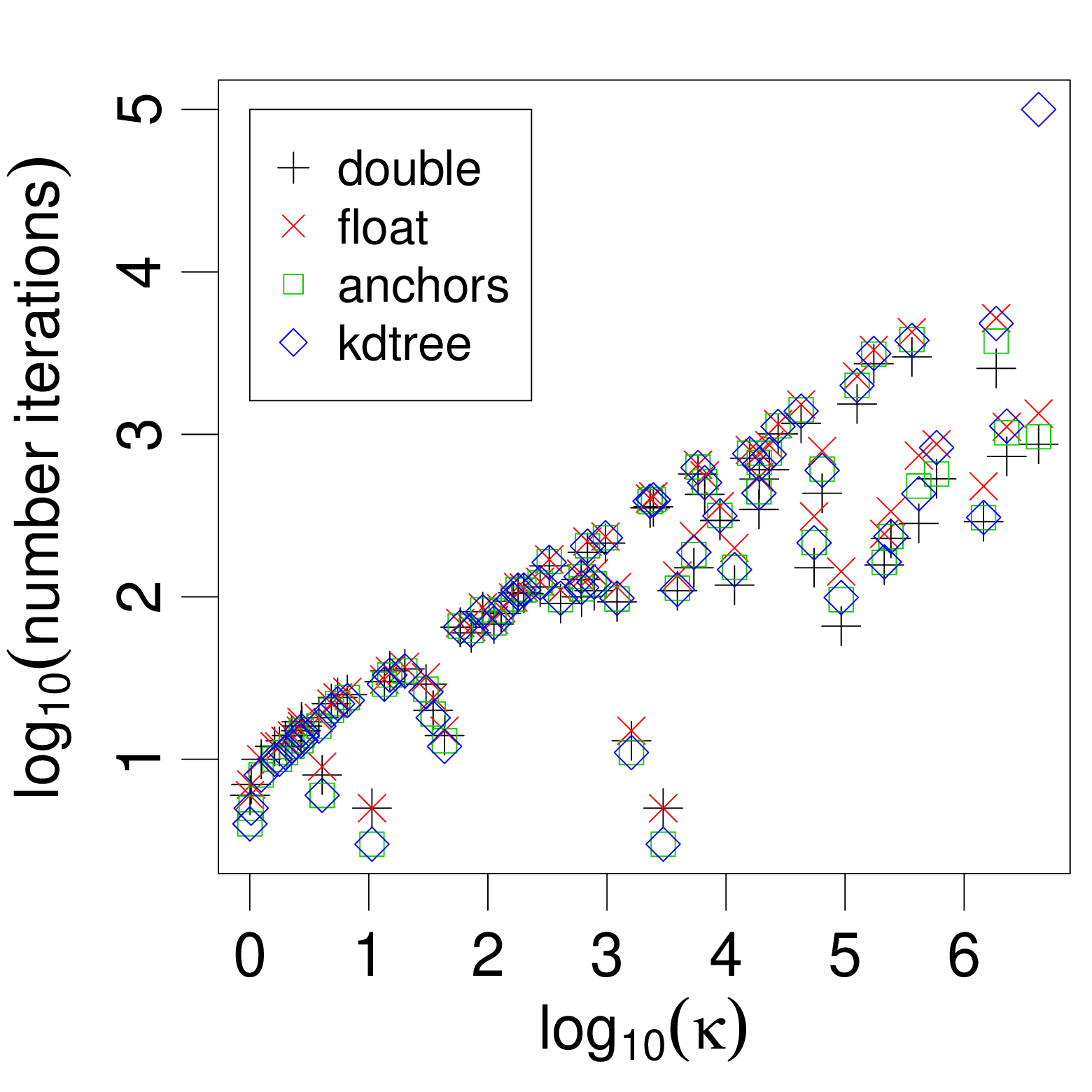} \includegraphics[width=0.49\columnwidth]{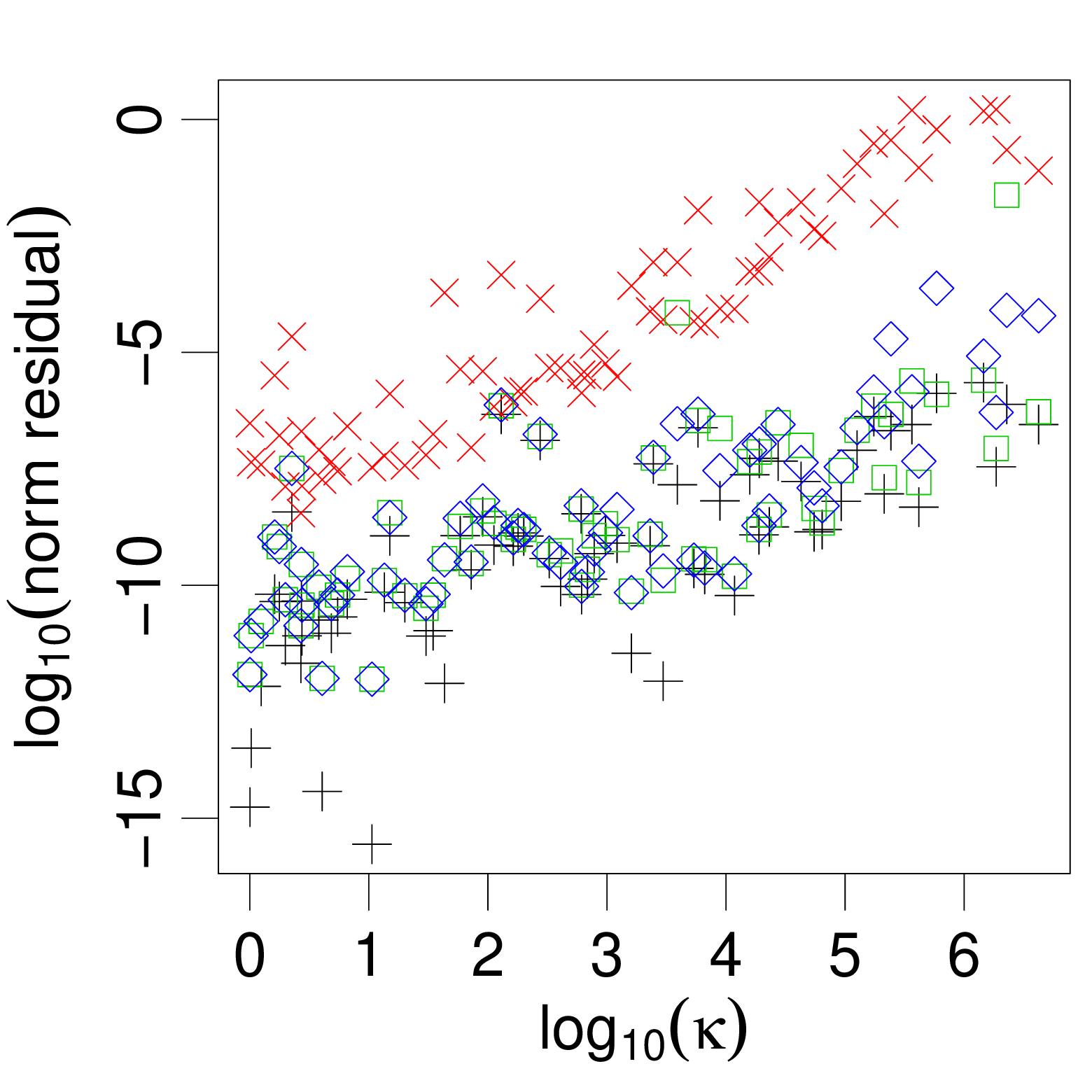}}
\centerline{\includegraphics[width=0.49\columnwidth]{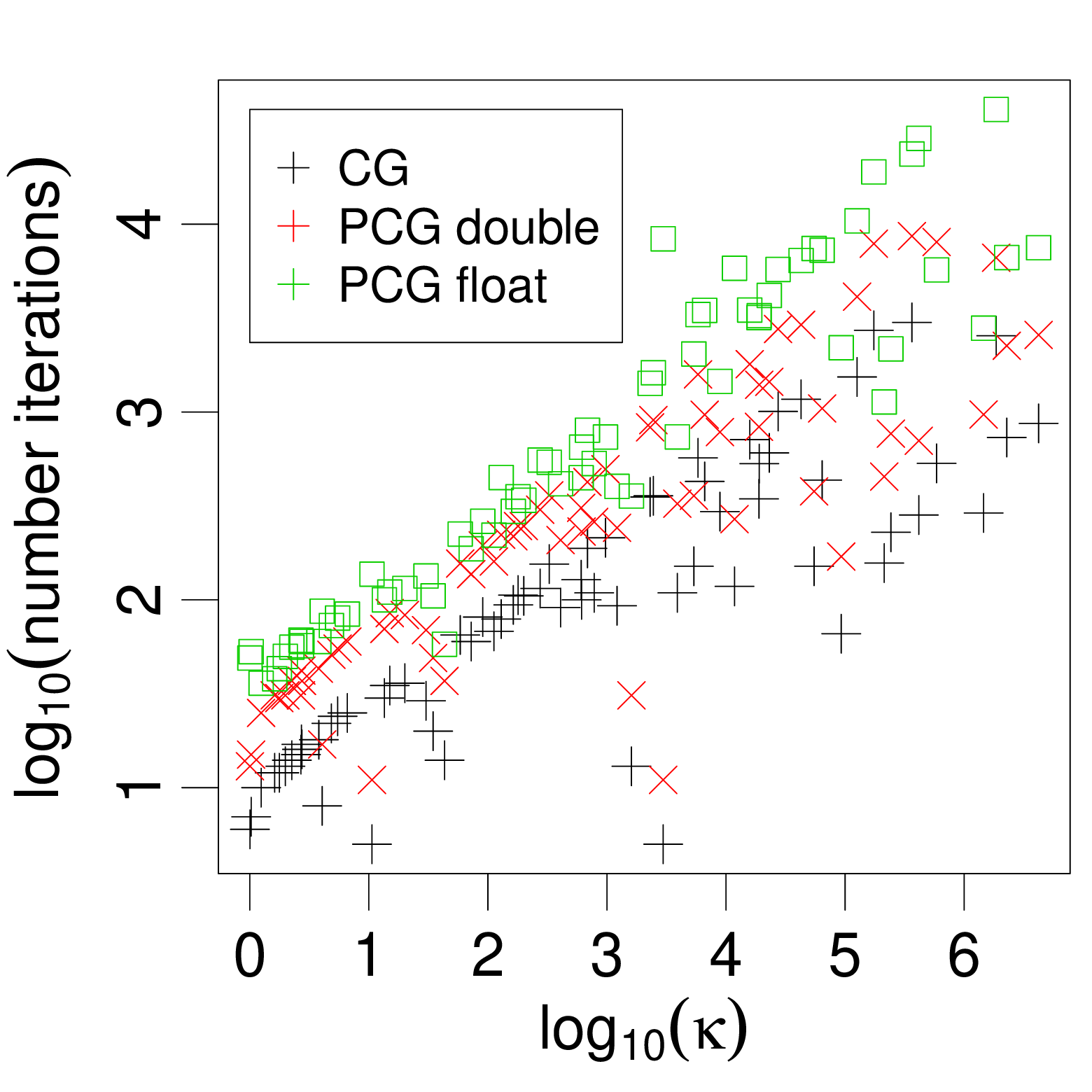} \includegraphics[width=0.49\columnwidth]{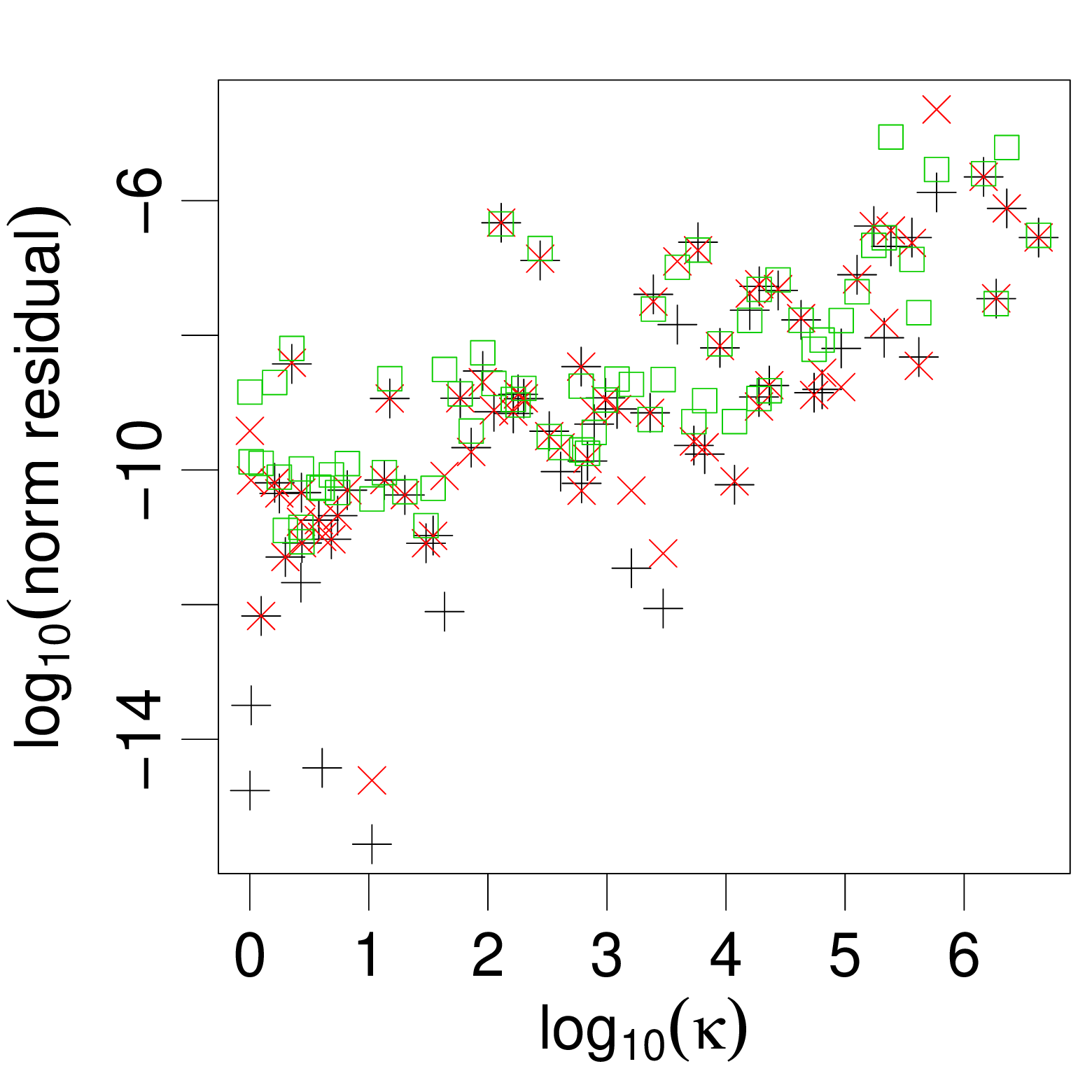}}
\caption{Top row: 
Comparison of fast CMVPs on number of iterations and error versus condition number.
Bottom row:
Comparison of the CG algorithm and two PCG algorithms using double and single precision CMVPs to solve inner linear systems.
}
\label{fig:compare:matrix:vector:preconditioning}
\end{center}
\vskip -0.2in
\end{figure}

In the top row of Fig.~\ref{fig:compare:matrix:vector:preconditioning} we show the number of iterations required by the CG algorithm to reach convergence versus the condition number and the error in the solution versus the condition number.
The error is defined as the norm of the difference between the solution obtained by the CG algorithm and the one obtained by factorizing $K$ using the Cholesky algorithm and carrying out forward and back substitutions with $\yvect$.
We compare a baseline CG algorithm with CMVPs performed in double precision with CG algorithms implemented with (i) single precision (``float'') CMVPs, (ii) ``kdtree'' CMVPs and (iii) ``anchors'' CMVPs.
The convergence threshold of the CG algorithm was set to $10^{-8}$, so in order to be able to satisfy this criterion when employing ``kdtree'' and ``anchors'' CMVPs, we selected the relative and absolute tolerance parameters to be $10^{-10}$.

%% There are a number of observations that we can make by analyzing these results.
%% The number of iterations generally grows with the condition number, sometimes well beyond $n$.
The results indicate that double precision calculations lead to the lowest number of iterations compared to the other methods, especially when $\kappa$ is large.
%% The error grows with the condition number too, and 
Double precision calculations also offer the lowest error.
Single precision calculations lead to a very poor error compared to the other methods.
The CG algorithm with ``kdtree'' CMVPs seems to take longer to converge than the one with ``anchor'' CMVPs, but it achieves a lower error. 

Drawing definitive conclusions on whether fast CMVPs yield any gain in computing time is far from trivial, as this very much depends on implementation details and hardware where the code is run.
What we can say, however, is that gaining orders of magnitude speed-ups would require reducing the accuracy of fast CMVPs, but this would require loosening up the convergence criterion in order for the CG algorithm to converge.
As a result, we would be able to obtain solutions of linear systems faster but at the cost of a reduced accuracy in the solution, which in turn would bias the estimation of gradients.

%% \begin{figure}[ht]
%% \vskip 0.2in
%% \begin{center}
%% \centerline{\includegraphics[width=\columnwidth]{figures/PLOT_ERROR_vs_CONDITION_NUMBER.eps}}
%% \caption{Caption.}
%% \label{fig:xxx}
%% \end{center}
%% \vskip -0.2in
%% \end{figure} 

%% \begin{figure}[ht]
%% \vskip 0.2in
%% \begin{center}
%% \caption{Caption.}
%% \label{fig:compare:preconditioning}
%% \end{center}
%% \vskip -0.2in
%% \end{figure} 

\subsection{Preconditioned CG}

The Preconditioned CG (PCG) is a variant of the CG algorithm that aims at mitigating the issues associated with the rate of convergence of the CG algorithm when the condition number $\kappa$ is large.
A (right) preconditioning matrix $J$ operates on the linear system yielding $$K J^{-1} (J \svect) = \bvect$$ %the CG algorithm is applied to the latter linear system.
The success of PCG is based on the possibility to construct $J$ so that $K J^{-1}$ is well conditioned.
This can be achieved when $J^{-1}$ well approximates $K^{-1}$, and a complication immediately arises on how to do so for general kernel matrices without carrying out expensive operations (in $O(n^3)$).

In \cite{Srinivasan14} it was proposed to define $J = K + \delta I$ with $\delta > 0$.
Compared to the standard CG algorithm, the PCG algorithm introduces the solution of an ``inner'' linear system of the form $J^{-1} \zvect$ at each iteration, that can be solved again using the CG algorithm.
%% The tradeoff between the number of iterations of the inner and outer CG algorithms is determined by the parameter $\delta$.
A large value of $\delta$ makes $K + \delta I$ well conditioned and makes convergence speed of the inner CG algorithm faster, whereas it makes $J^{-1}$ and $K^{-1}$ considerably different leading to the necessity to run the outer CG algorithm for several iterations.
For small values of $\delta$ the situation is reversed, so $\delta$ needs to be tuned to find an optimal compromise. %tradeoff between convergence speed in the inner and outer iterations need to be found by tuning $\delta$ properly.

In the bottom row of Fig.~\ref{fig:compare:matrix:vector:preconditioning}, we compare the standard CG algorithm with two versions of the PCG algorithm on number of iterations and accuracy of the solution.
In the first version of the PCG algorithm we used double precision calculations when solving the inner linear systems, whereas in the second version we used single precision calculations.
In both versions of the PCG algorithm we set $\delta$ to yield the lowest number of iterations in order to show whether it is possible to reduce the number of computations.

The results show that the standard CG algorithm takes less iterations to converge than the PCG algorithm (counting both inner and outer iterations).
Even in the case of single precision calculations in the inner CG algorithm, we did not experience any gain in computing time due to the increased number of iterations. 
%% This is somewhat in contrast with what reported i
%% This indicates that there is some degree of sensitivity of the results across different datasets, covariance structure, and experimental conditions (
%% although experimental conditions are different as
For other data and in different experimental conditions there might be a computational advantage in using a preconditioner, as shown in \cite{Srinivasan14}, but the gain is generally modest. % suggest that there are indeed cases where there is a computational advantage in using a preconditioner. %; the differences are that in their work they use different data, $\kappa$ is increased by varying $\tau$ and $\sigma = 1$ in the RBF kernel in eq.~\ref{eq:covariance:rbf}, and a looser convergence criterion in the inner and outer CG algorithms is used.
%% ).
%% In the case of single precision calculations for the inner CG algorithm, a higher number of iterations could be counterbalanced by the fact that CMVPs are much faster in single precision.
%% If single precision CMVPs required $1/4$ of the time taken by double precision ones, we would improve speed using float in $25\%$ of the cases reported in the figure.
%% In our case, single precision CMVPs take $1/3$ of the time of the running time compared to double precision and this results in no gains in computing time.
%% These results are somewhat in contrast to what reported in \cite{Srinivasan14} where it is showed that the PCG algorithm can lead to a reduction in computing time.

\section{Unbiased LInear System SolvEr (ULISSE)} \label{sec:ulisse}

From the analysis in the previous sections it is evident that none of the standard ways to speedup calculations and convergence of the CG algorithm offer substantial gains in computing time.
As a result, employing iterative methods as an alternative to traditional factorization techniques seems beyond practicality as pointed out, e.g., in \cite{Murray09}.
One of the novel contributions of this paper is to accelerate the CG algorithm at the expenses of obtaining an (unbiased) estimate of the solution. % of linear systems rather than computing it exactly.%  to yield unbiased solutions of linear systems, while being orders of magnitude faster than solving linear systems exactly.
The idea is to stop the CG algorithm before the convergence criterion is satisfied and apply some corrections to ensure unbiasedness of the solution.
We note here that our proposal can be applied to any of the variants of the CG algorithm presented earlier and to dense as well as sparse linear systems.
%% We point out here that this idea is quite general as it can be applied regardless of the particular choice of kernel vector multiplication and in the PCG algorithm as well.

%% The analysis of convergence of the CG algorithm shows that the iterative updates will approach the solution almost surely \cite{Golub96}, so at first it is not trivial to see how to stop the iterations early in a way that yields an unbiased solution.
%% This is in contrast with the analysis of convergence of the steepest descent algorithm that is based on the observation that residuals approach a zero mean Gaussian distribution with variance that shrinks to zero as iterations go by \cite{Golub96}.
%% Steepest descent would therefore be amenable to early stop preserving unbiasedness of the solution, but in the case of full covariance matrices convergence would be extremely slow and therefore impractical.

We can rewrite the final solution of a linear system obtained by the CG algorithm as a sum of incremental updates
\begin{equation}
\svect = \svect_0 + \deltavect_1 + \ldots + \deltavect_T
\end{equation}
assuming that it takes $T$ iterations to satisfy the convergence threshold $\epsilon$.
We can define an ``early stop'' threshold $\alpha > \epsilon$ that will be reached after $l < T$ iterations, and rewrite the final solution by introducing a series of coefficients as follows
%% We can define an early stop threshold $\alpha$ that is larger than the convergence threshold $\epsilon$ after $l$ iterations.
%% What we can do, is to 
\begin{eqnarray}
\svect & = & \svect_0 + \sum_{i=1}^{l-1} \deltavect_i + \frac{1}{w_0} \biggl( w_0 \deltavect_{l+0} + \frac{1}{w_1} \biggl( w_0 w_1 \deltavect_{l+1} + \nonumber \\
& & + \frac{1}{w_2} \left(w_0 w_1 w_2 \deltavect_{l+2} + \ldots \right) \biggr) \biggr)
\end{eqnarray}
We will focus on coefficients defined as $w_r = \exp(\beta r)$, but this choice is not restrictive.
%% This choice ensures that any $\frac{1}{w_r} \leq 1$.
We can now obtain an unbiased estimate of the solution of the linear system by adding these instructions to the standard CG algorithm: set $\tilde{\svect} = \svect_0 + \sum_{i=1}^{l-1} \deltavect_i$ and iterate for $j = 0, 1, \ldots$ the following two steps (i) draw $u_j \sim U[0,1]$ (ii) if $u_j < \frac{1}{w_j}$ then $\tilde{\svect} = \tilde{\svect} + \prod_{r=0}^j w_r \deltavect_{l+j}$, else return $\tilde{\svect}$ and stop the CG algorithm.
The expectation of $\tilde{\svect}$ is clearly $\svect$ and the rate of decay $\beta$ in the expression of $w_r$ determines the average number of steps that are carried out after the convergence threshold $\alpha$ is reached.
%% The accumulation of the terms forming $\tilde{\svect}$ is carried out during the execution of the standard CG algorithm.
%% These terms contain a potentially large factor $\prod_{r=0}^j w_r$, so the threshold $\alpha$ should be chosen to ensure that the values of $\deltavect_{l+j}$ counterbalance them.
%% In practice we simulate the acceptance steps beforehand and know exactly for how many iterations run the CG algorithm after early stop.
%% We remark that for this to work we need to continue running the standard CG algorithm, and accumulate the results in $\tilde{\svect}$ once the algorithm reached early convergence.
%% The vector $\tilde{\svect}$ is never fed back to the CG algorithm, as this would not lead to an unbiased estimator for $\svect$; this is easily seen by analyzing the complex relationship between ...

%% A simple analysis of the expansion coefficients shows that $j$ iterations after early stop we need to compute the product
%% \begin{equation}
%% \prod_{r=0}^j w_r % = \exp{\beta \sum_{r=0}^j r} 
%% = \exp\left(\beta \frac{j (j+1)}{2}\right)
%% \end{equation}
%% and that the expected number of steps taken after early stop is $\sum_{j=0}^{\infty} \prod_{r=0}^j \frac{1}{w_r}$ which is roughly $1.4$ for $\beta=1$, $4.0$ for $\beta=10$, and $12.6$ for $\beta=100$.

\begin{figure}[t]
% \vskip 0.2in
\begin{center}
\centerline{\includegraphics[width=0.49\columnwidth]{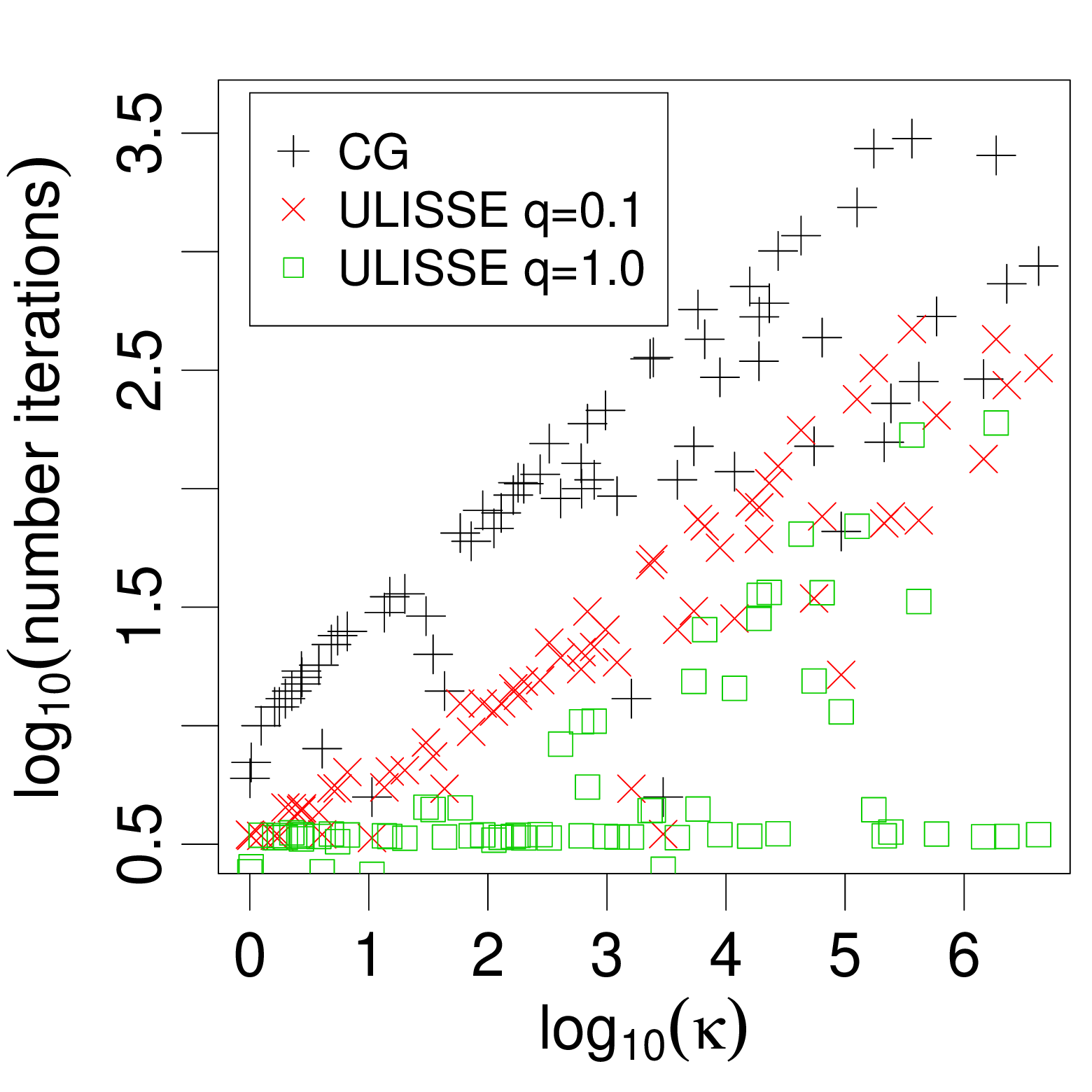} \includegraphics[width=0.49\columnwidth]{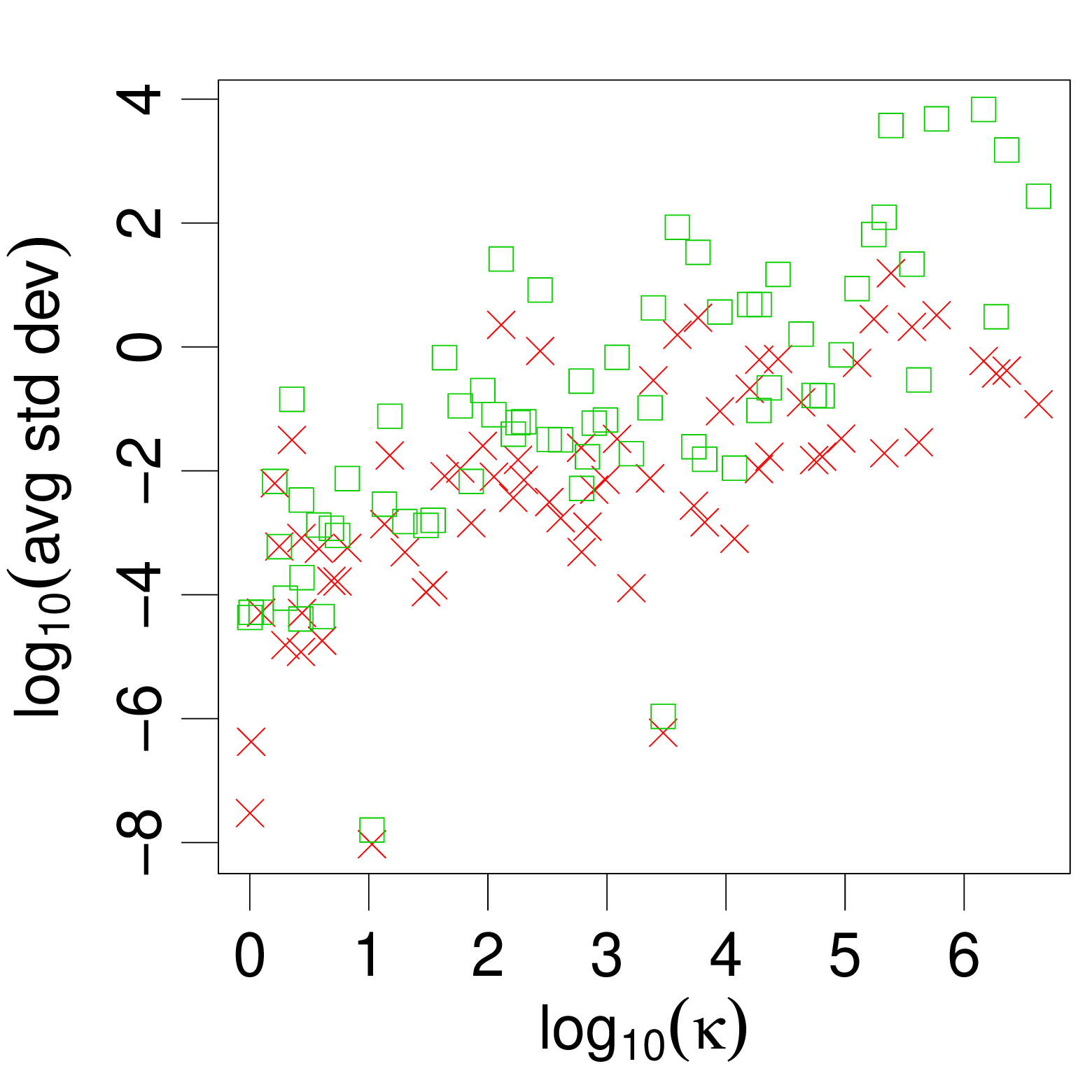}}
%% \caption{$\gamma = 1$.}
%% \label{fig:xxx}
%% \end{center}
%% \vskip -0.2in
%% \end{figure} 
%% \begin{figure}[ht]
%% \vskip 0.2in
%% \begin{center}
\centerline{\includegraphics[width=0.49\columnwidth]{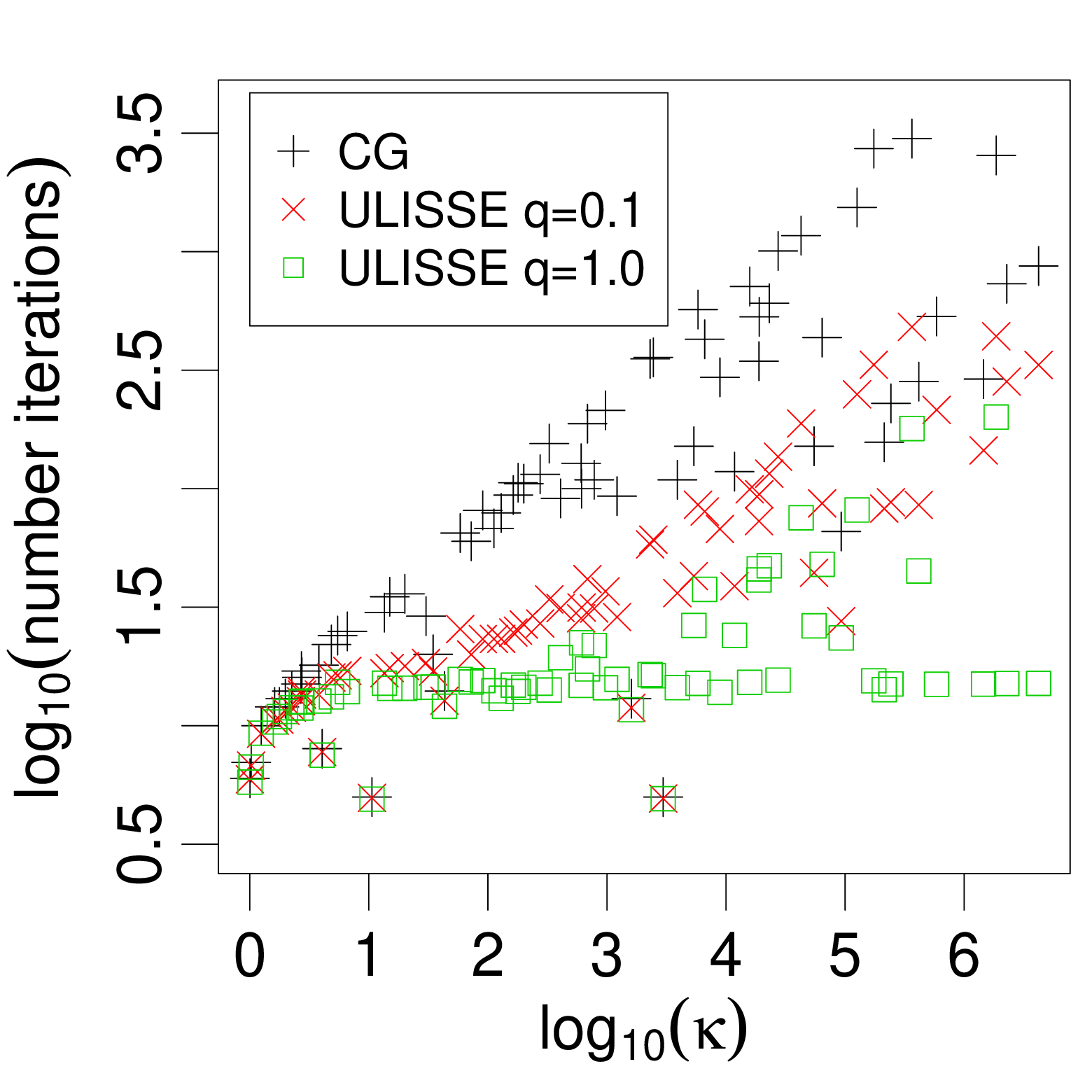} \includegraphics[width=0.49\columnwidth]{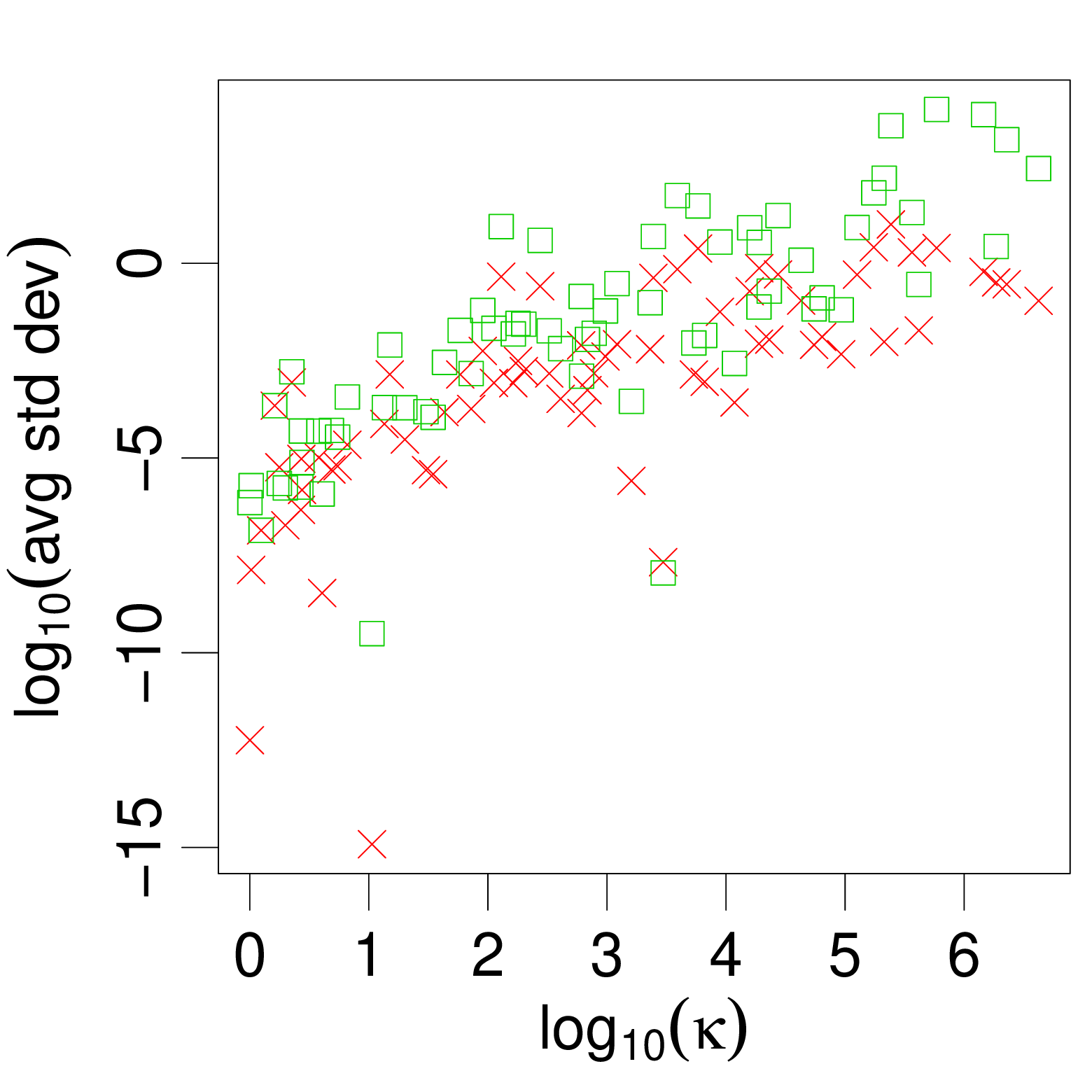}}
\caption{Comparison of the CG algorithm and ULISSE with early stop thresholds $\alpha$ calculated with $q = 0.1$ and $q = 1$ on number of iterations and standard deviation of the solution. The top row corresponds to $\beta = 1$ in the calculation of the weights $w_r$, whereas the bottom row corresponds to $\beta = 100$.}
\label{fig:compare:early:stop}
\end{center}
\vskip -0.2in
\end{figure} 

For simplicity, we set the early stop threshold to $\alpha = q \sqrt{n}$ as $q$ gives a rough indication of the average error that we are expecting in each element of the solution.
In Fig.~\ref{fig:compare:early:stop} we report number of iterations and average standard deviation across the elements of the solution.
ULISSE with two different values of $\beta$ and $q$ is compared with the baseline CG algorithm without early stop (``CG'').
We stress again that the error is such that the solution is unbiased.

\subsection{Impact on the calculation of stochastic gradients}

We conclude this section by showing the impact of ULISSE in the calculation of stochastic gradients in GPs.
Applying the proposed unbiased solver to the first term of $\tilde{g}_i$ in eq.~\ref{eq:gradient:unbiased} is straightforward and it requires solving $N_{\rvect}$ linear systems, one for each of the $\rvect^{(i)}$ vectors.
For the quadratic term in $\yvect$, instead, we need to obtain two independent unbiased estimates of $K^{-1} \yvect$ in order for the expectation of the whole term to be unbiased.
This can be implemented by running a single instance of the CG algorithm and keeping track of two solutions obtained by independent draws of the uniform variables $u_j$ used to early stop the CG algorithm.
We remark that the unbiased estimation of gradients involves now two sources of stochasticity: one due to the stochastic estimate of the trace term in eq.~\ref{eq:gradient:exact}, and one due to the proposed way to unbiasedly solve all linear systems in eq.~\ref{eq:gradient:unbiased}.
%% \begin{equation}
%% \tilde{g}_i = 
%% - \frac{1}{2} \rvect^{\T} K^{-1} \frac{\partial K}{\partial \theta_i} \rvect + \yvect^{\T} K^{-1} \frac{\partial K}{\partial \theta_i} K^{-1} \yvect
%% \end{equation}

\begin{figure}[t]
% \vskip 0.2in
\begin{center}
\centerline{\includegraphics[width=0.49\columnwidth]{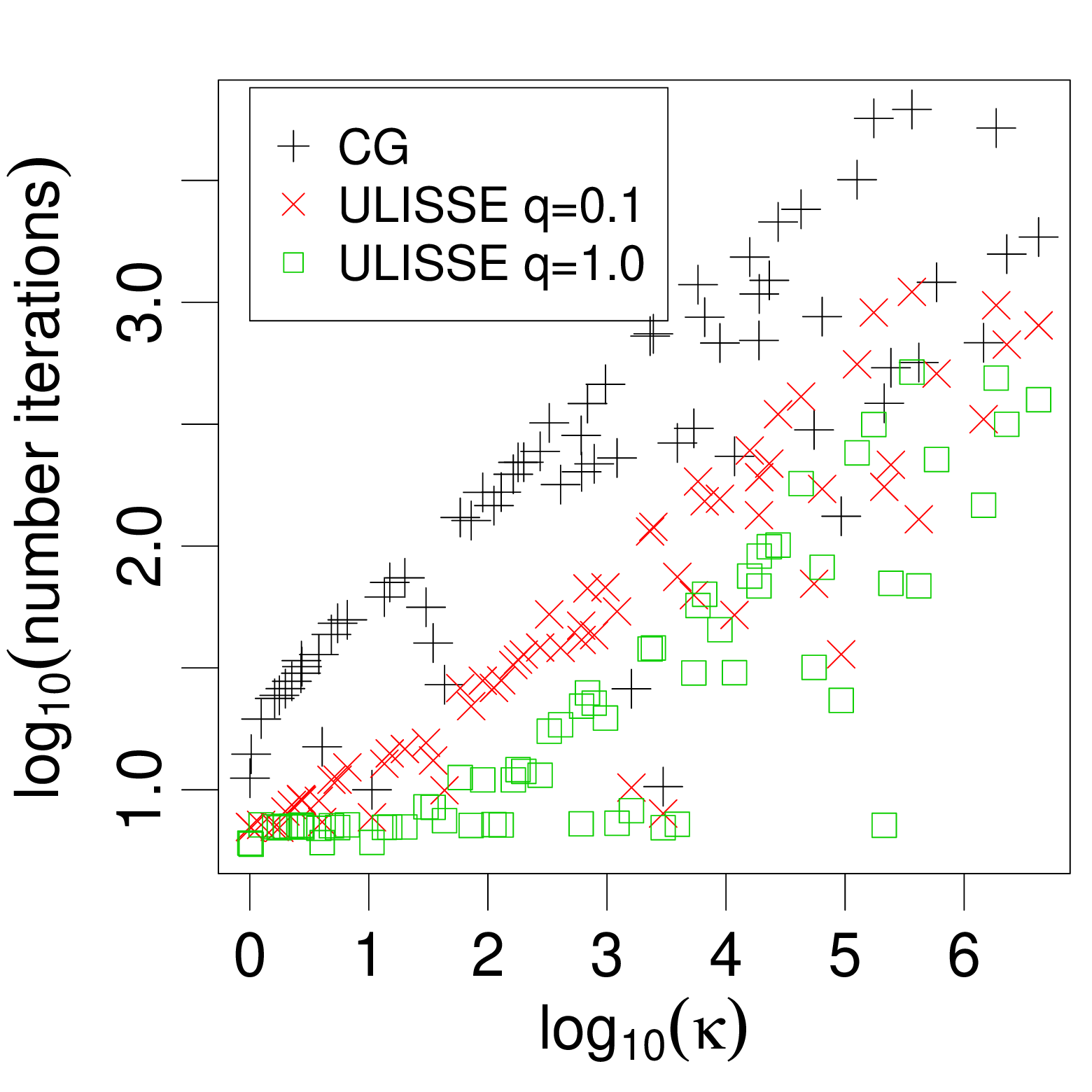} \includegraphics[width=0.49\columnwidth]{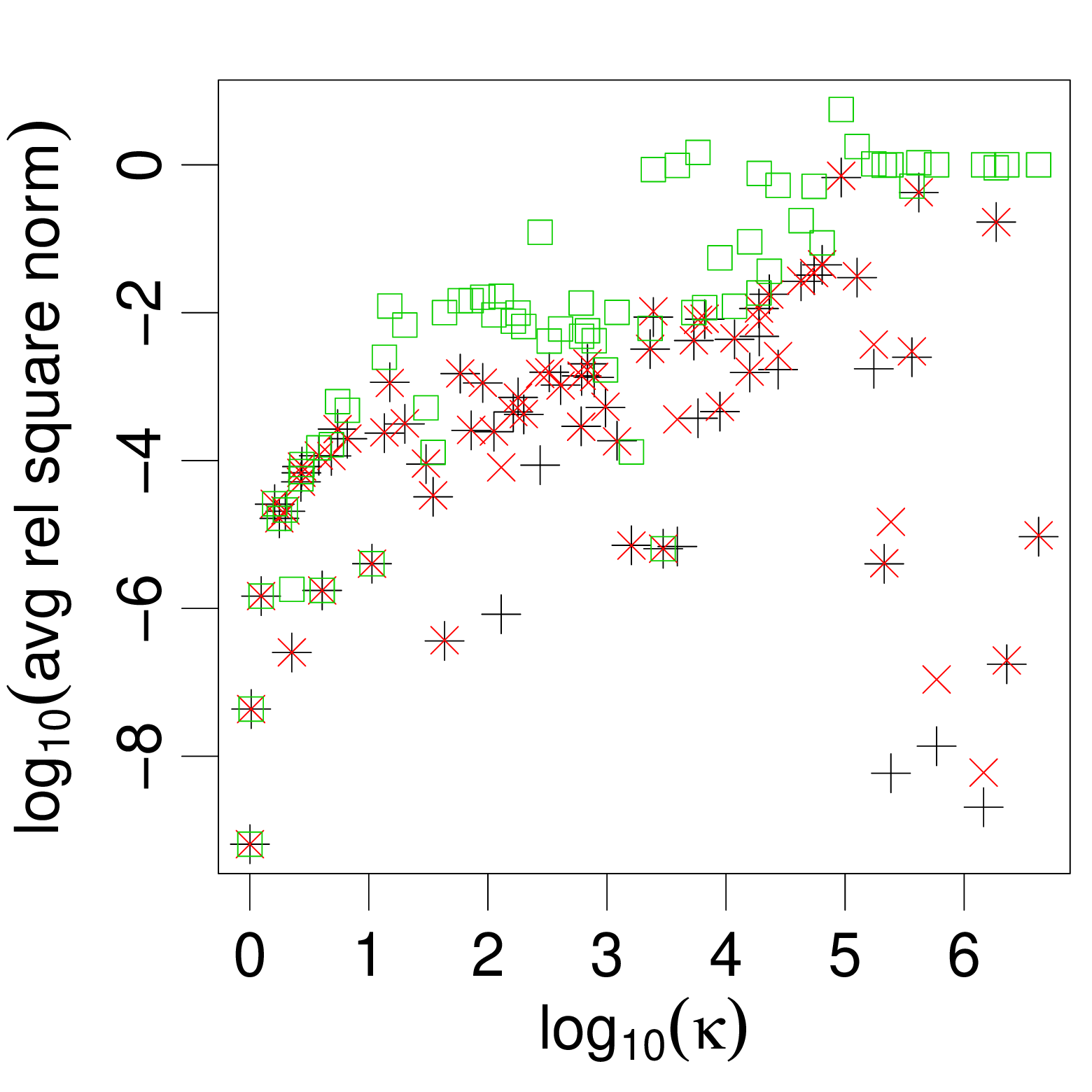}}
%% \caption{$\gamma = 1$.}
%% \label{fig:xxx}
%% \end{center}
%% \vskip -0.2in
%% \end{figure} 
%% \begin{figure}[ht]
%% \vskip 0.2in
%% \begin{center}
%% \centerline{\includegraphics[width=0.49\columnwidth]{} \includegraphics[width=0.49\columnwidth]{figures/PLOT_EARLYSTOP_GAMMA_100_AVG_STDEV_vs_CONDITION_NUMBER.eps}}
\caption{Comparison of the CG algorithm and ULISSE and early stop thresholds computed with $q=0.1$ and $q=1$ to estimate the gradient of the log-marginal likelihood in eq.~\ref{eq:gradient:unbiased}. In ULISSE, the weights $w_r$ are calculated with $\beta=1$.}
\label{fig:comparison:gradient}
\end{center}
\vskip -0.2in
\end{figure} 

Fig.~\ref{fig:comparison:gradient} reports the average, taken with respect to $100$ repetition of the $\log_{10}$ of the relative square norm of the error:
\begin{equation}
\frac{\| \mathbf{g}(\thetavect) - \tilde{\mathbf{g}}(\thetavect) \|^2}{\| \mathbf{g}(\thetavect) \|^2}
\end{equation}
as a function of the condition number $\kappa$.
We used one vector $\rvect^{(1)}$ to estimate the gradient in eq.~\ref{eq:gradient:unbiased}.
%% Again, we stress that the estimates are unbiased.
The figure shows that the estimate in eq.~\ref{eq:gradient:unbiased} (``CG'' in the figure) is quite accurate, as the relative error is small in a wide range of values of $\kappa$.
Also, at the expenses of a larger variance in the estimate of the gradient, ULISSE yields orders of magnitude improvements in the number of iterations.
%% This is the key aspect that will make it possible to scale inference for GPs as demonstrated in the next section.

\section{Experimental validation} \label{sec:experiments}

In this section, we infer covariance parameters of GP regression models using SGLD with ULISSE.
We start by considering the Concrete data set where it is possible to compare our proposal with the Metropolis-Hastings (MH) algorithm.
%% In this way, it is possible to gain some insights into the behavior of the proposed algorithm and check that we can actually employ the SGLD algorithm to infer parameters in GP models.
We then demonstrate the scalability of the proposed methodology by considering a data set with $n = 22,784$ and $d=8$.
%% The census dataset, which comprises $n = 22,784$ input vectors in $d = 8$ dimensions and where it is not possible to run any of the standard 
%% We aim to show the feasibility of the proposed method to infer covariance parameters for GPs applied to a dataset of this size.
%% We are not aware of any methods to carry out Bayesian inference for data of this size without employing deterministic approximations.

\subsection{Comparison with MCMC}

%% \begin{figure*}[t]
%% \vskip 0.2in
%% \begin{center}
%% \centerline{\includegraphics[width=5.5cm]{figures/PLOT_COMPARE_SGR_MCMC_psi_sigma.eps} \includegraphics[width=5.5cm]{figures/PLOT_COMPARE_SGR_MCMC_psi_lambda.eps} \includegraphics[width=5.5cm]{figures/PLOT_COMPARE_SGR_MCMC_psi_tau.eps} }
%% \centerline{\includegraphics[width=5.5cm]{figures/PLOT_PSRF_CONCRETE_psi_sigma.eps} \includegraphics[width=5.5cm]{figures/PLOT_PSRF_CONCRETE_psi_lambda.eps} \includegraphics[width=5.5cm]{figures/PLOT_PSRF_CONCRETE_psi_tau.eps} }
%% \caption{Caption.}
%% \label{fig:compare:mcmc:concrete}
%% \end{center}
%% \vskip -0.2in
%% \end{figure*} 

\begin{figure}[t]
% \vskip 0.2in
\begin{center}
\centerline{\includegraphics[width=0.49\columnwidth]{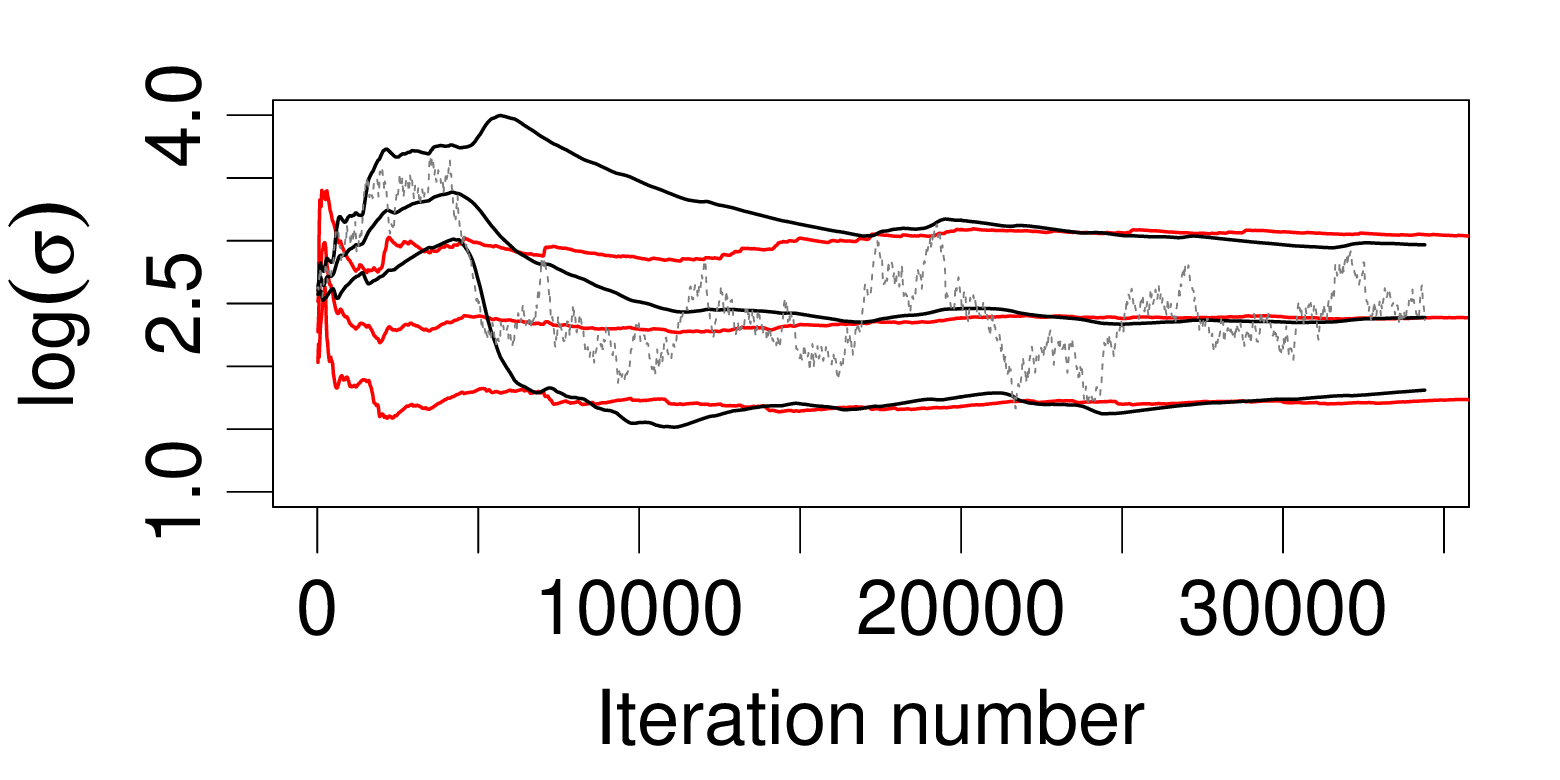} \includegraphics[width=0.49\columnwidth]{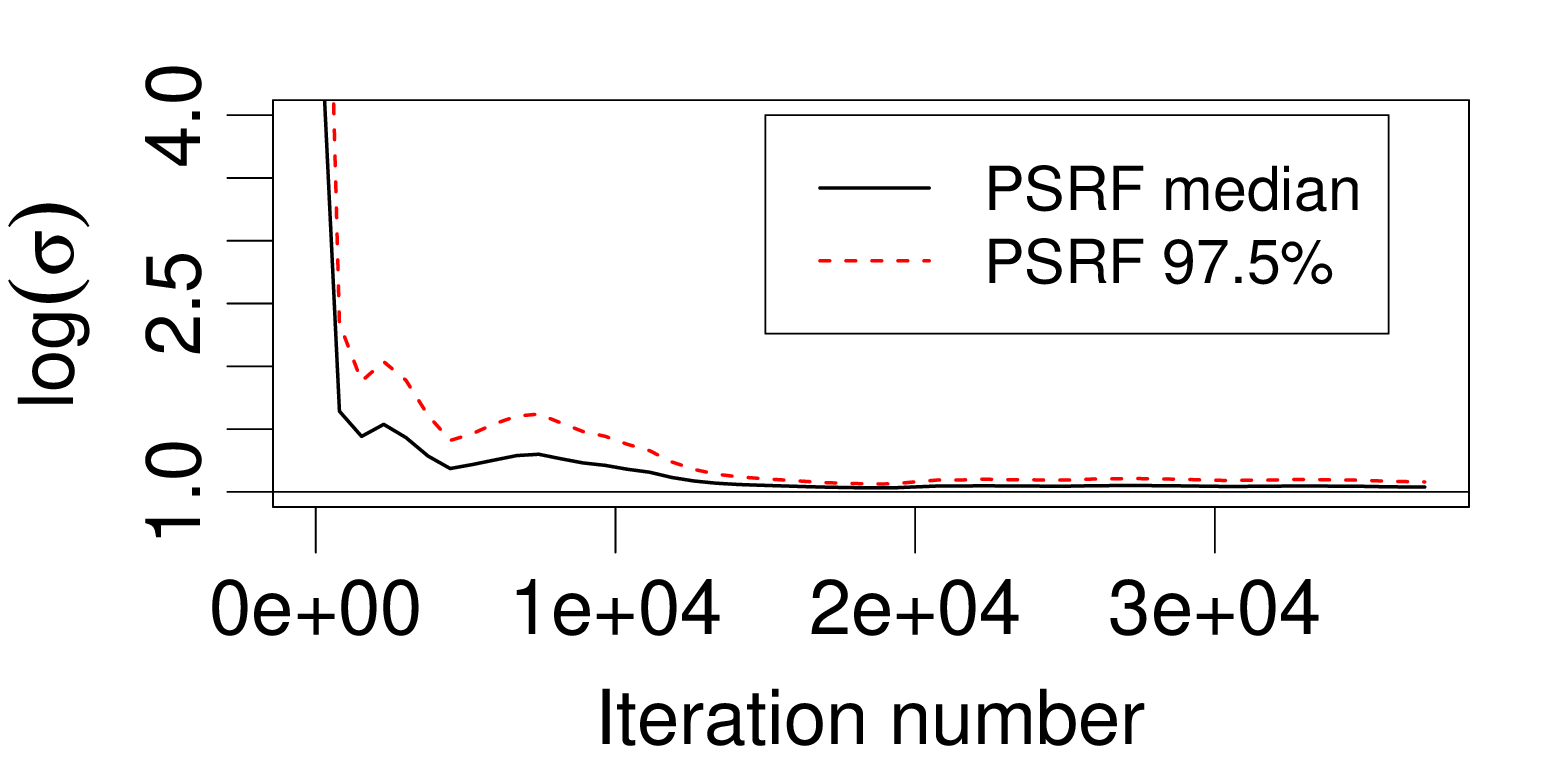}}
\centerline{\includegraphics[width=0.49\columnwidth]{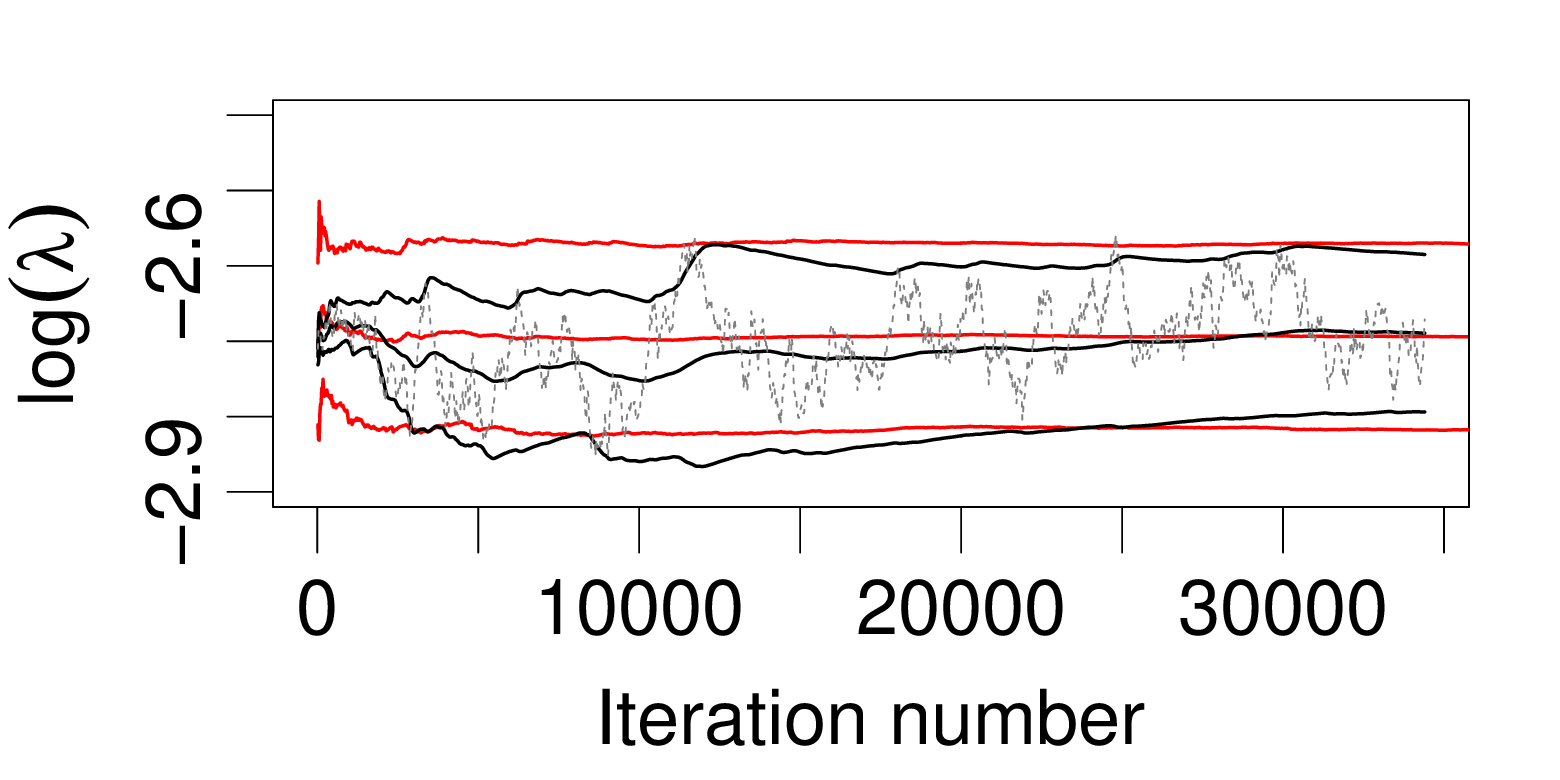} \includegraphics[width=0.49\columnwidth]{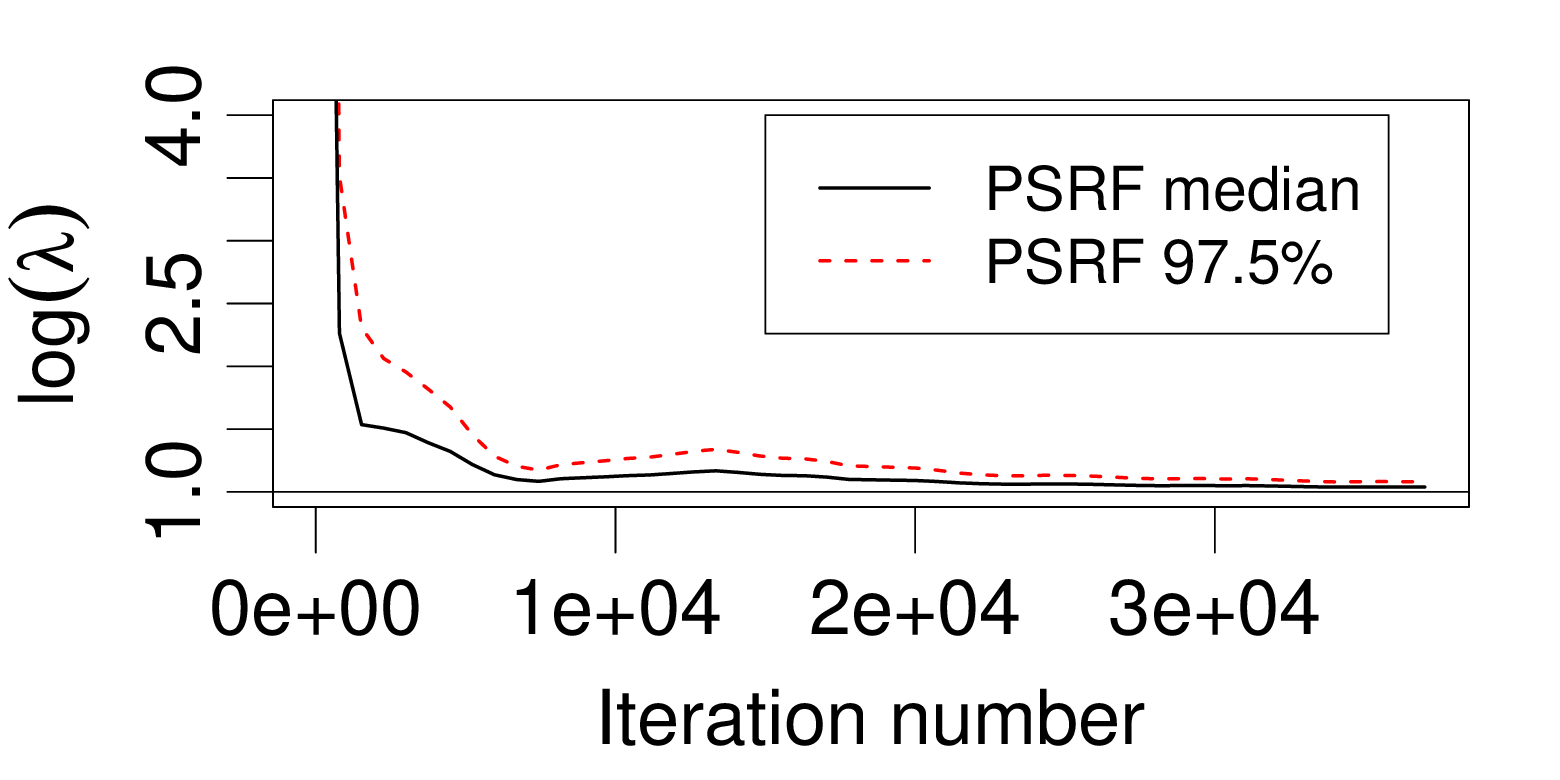}}
\centerline{\includegraphics[width=0.49\columnwidth]{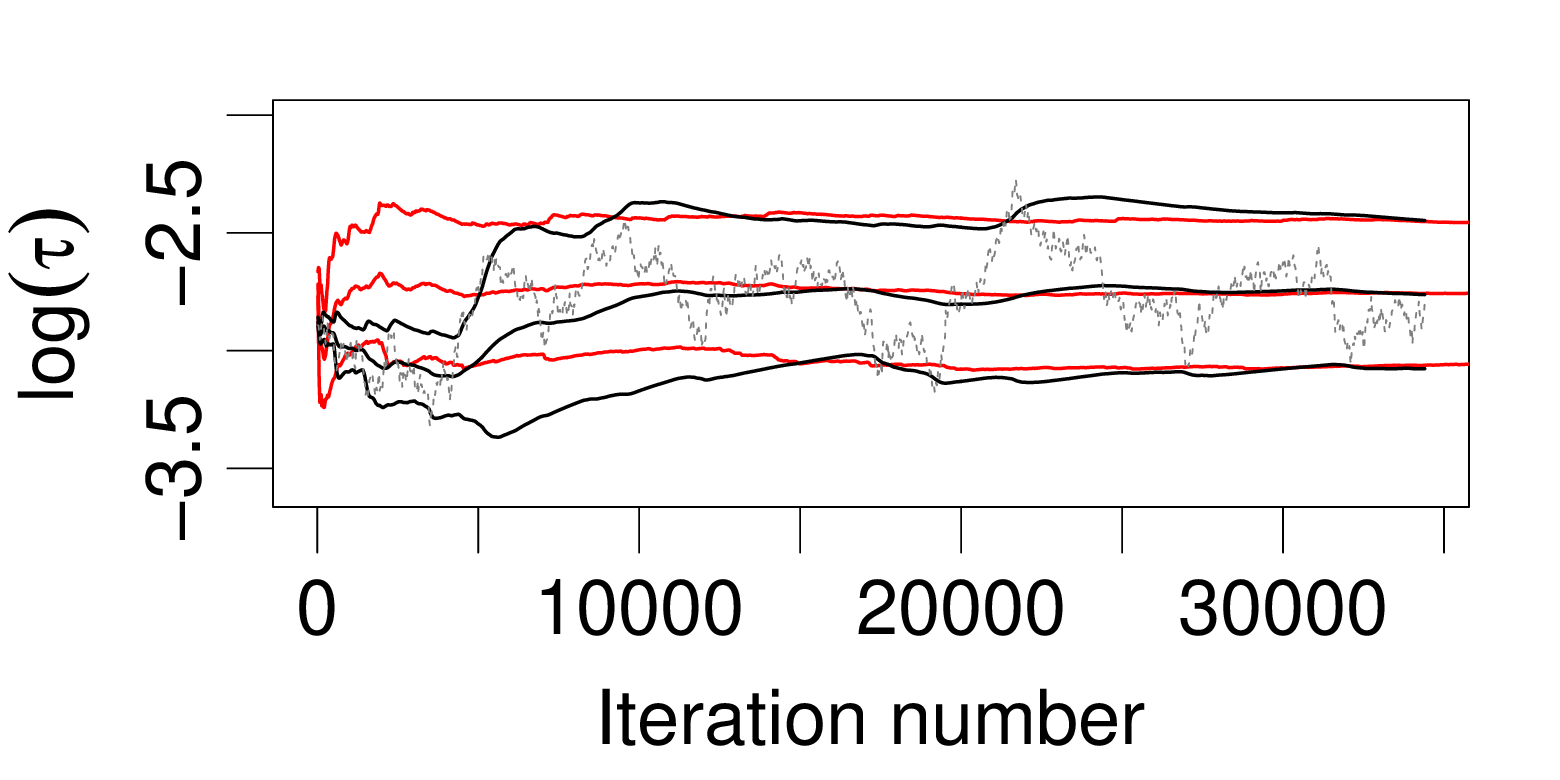} \includegraphics[width=0.49\columnwidth]{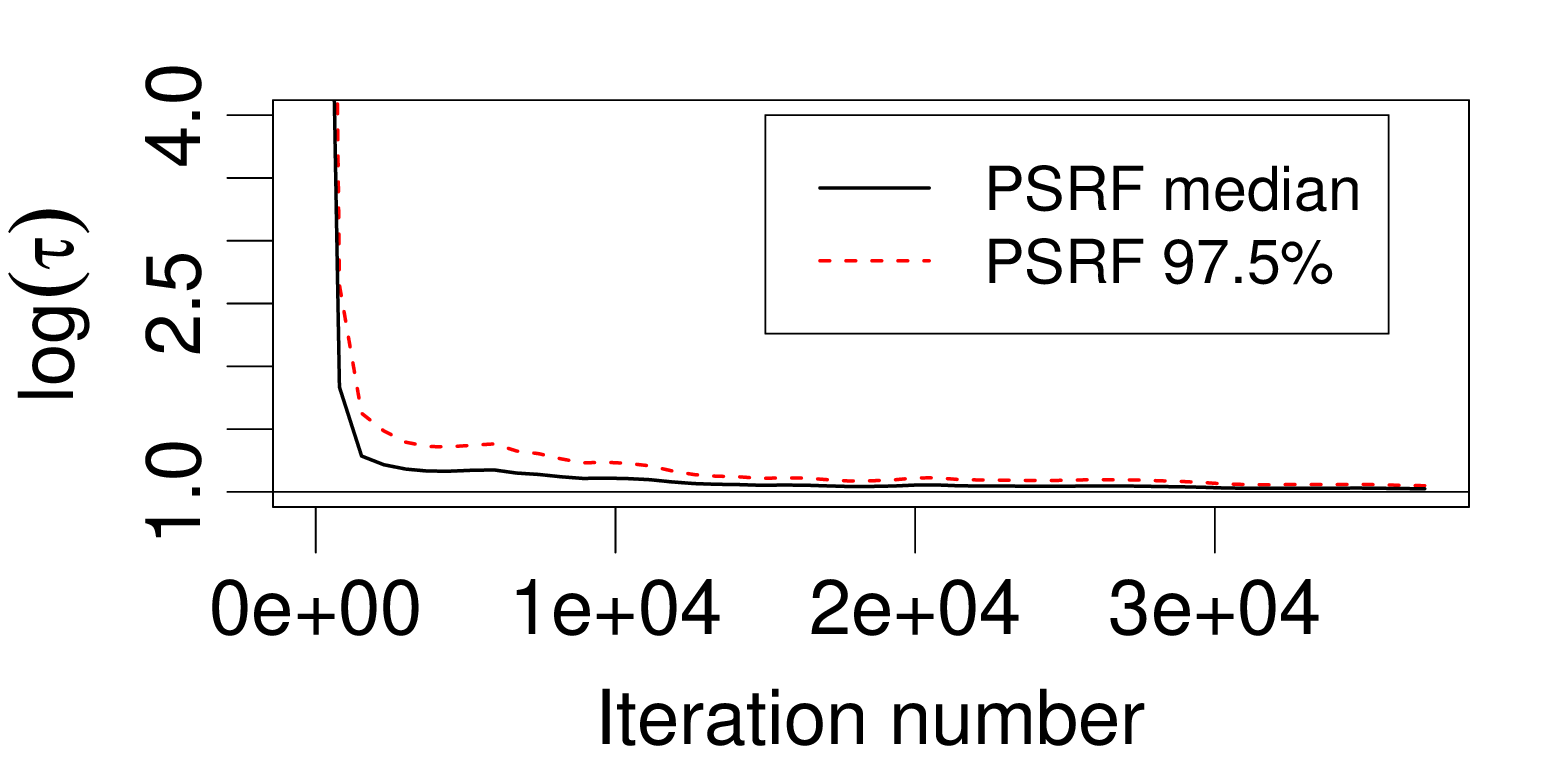}}
\caption{Concrete data ($n=1030$) - Left panel: Comparison of MCMC (red) and SGLD with ULISSE (black) on running mean and plus/minus two standard deviations. The trace of one chain of SGLD is shown in gray. Right panel: Convergence analysis of SGLD with ULISSE with PSRF computed over ten chains.}
\label{fig:compare:mcmc:concrete}
\end{center}
\vskip -0.2in
\end{figure}

We ran the MH algorithm for fifty-thousand iterations to the GP regression model with covariance in eq.~\ref{eq:covariance:rbf} applied to the Concrete data set. % with covariance as in eq.~\ref{eq:covariance:rbf}. 
We allowed for an initial adaptive phase to reach an average acceptance rate between $0.2$ and $0.4$, and we discarded the first ten-thousand samples.
%% We used the remaining samples to estimate the mean and standard deviation of the posterior distribution over the three parameters.
Fig.~\ref{fig:compare:mcmc:concrete} shows the running mean and the interval corresponding to plus/minus twice the running standard deviation of the posterior over the three parameters (solid red lines) computed over the remaining forty-thousand samples.

We compare the run from the MH algorithm with SGLD, where we made the following design choices.
We employed ULISSE within the CG algorithm with double precision CMVPs.
We set the early stop threshold $\alpha$ to $\sqrt{n}$ and the parameter $\beta$ in the computation of the weights $w_r$ to $1$.
Stochastic gradients were computed using $N_{\rvect} = 4$ vectors $\rvect^{(i)}$.
We ran SGLD for forty-thousand iterations; the step-size was set to $\varepsilon_t = a(b + t)^{-\gamma}$, with $\gamma = 1$, and it was chosen to start from $10^{-1}$ and reduce to $10^{-4}$ on the last iteration.
During the execution of SGLD we monitored the quantity 
$\frac{\varepsilon_t}{4} \lambda_{\max}\left( M^{\frac{1}{2}} V M^{\frac{1}{2}}\right)$
as discussed in Section~\ref{sec:ulisse}, and we froze the value of $\varepsilon_t$ when it was less than $0.002$; the covariance of the gradients $V$ was estimated on batches of one-hundred iterations.
In order to speed up computations, we decided to redraw the vectors $\rvect^{(i)}$ every twenty iterations and to keep them fixed in between.
The advantage of this is that the solutions of the linear systems $K \rvect^{(i)}$ can be used to initialize the same systems when proposing new $\thetavect$'s thus speeding up convergence.
Finally, we set the preconditioning matrix $M$ in SGLD as the inverse of the negative Hessian of the log of the posterior density at its mode computed on a subset of five hundred input vectors, as this is cheap way to obtain a rough idea of the covariance structure of the posterior distribution for the full data set.

SGLD yields an effective sample size of about $0.1\%$ and it draws one independent sample every $27$~sec.
In Fig.~\ref{fig:compare:mcmc:concrete} we report the running statistics for the three parameters (solid black lines), and the trace-plot of one run of SGLD (solid gray lines), where we discarded all iterations prior to the freezing of the step-size $\varepsilon_t$.
The figure shows a striking match between the results obtained by a standard MCMC approach and SGLD with ULISSE.
This demonstrates that our proposal is a valid alternative to other MCMC approaches to reliably quantify uncertainty in GPs.

In order to check convergence speed of SGLD, we ran ten parallel chains and computed the Potential Scale Reduction Factor (PSRF) \cite{Gelman92}.
The chains were initialized by drawing from a Gaussian with mean on the MAP solution over a subset of five hundred input vectors and covariance $M$, so as to ensure enough dispersion to reliably report the PSRF. 
Fig.~\ref{fig:compare:mcmc:concrete} shows the median and the $97.5$th percentile of the PSRF across the ten chains.
The analysis of these plots reveals that SGLD achieves convergence after few thousand iterations.

\subsection{Demonstration on a larger dataset}

%% \begin{figure*}[t]
%% \vskip 0.2in
%% \begin{center}
%% \centerline{\includegraphics[width=5.5cm]{figures/PLOT_CENSUS_SGR_psi_sigma.eps} \includegraphics[width=5.5cm]{figures/PLOT_CENSUS_SGR_psi_lambda.eps} \includegraphics[width=5.5cm]{figures/PLOT_CENSUS_SGR_psi_tau.eps} }
%% \centerline{\includegraphics[width=5.5cm]{figures/PLOT_PSRF_CENSUS_psi_sigma.eps} \includegraphics[width=5.5cm]{figures/PLOT_PSRF_CENSUS_psi_lambda.eps} \includegraphics[width=5.5cm]{figures/PLOT_PSRF_CENSUS_psi_tau.eps} }
%% \caption{Caption.}
%% \label{fig:final:census}
%% \end{center}
%% \vskip -0.2in
%% \end{figure*} 

\begin{figure}[t]
% \vskip 0.2in
\begin{center}
\centerline{\includegraphics[width=0.49\columnwidth]{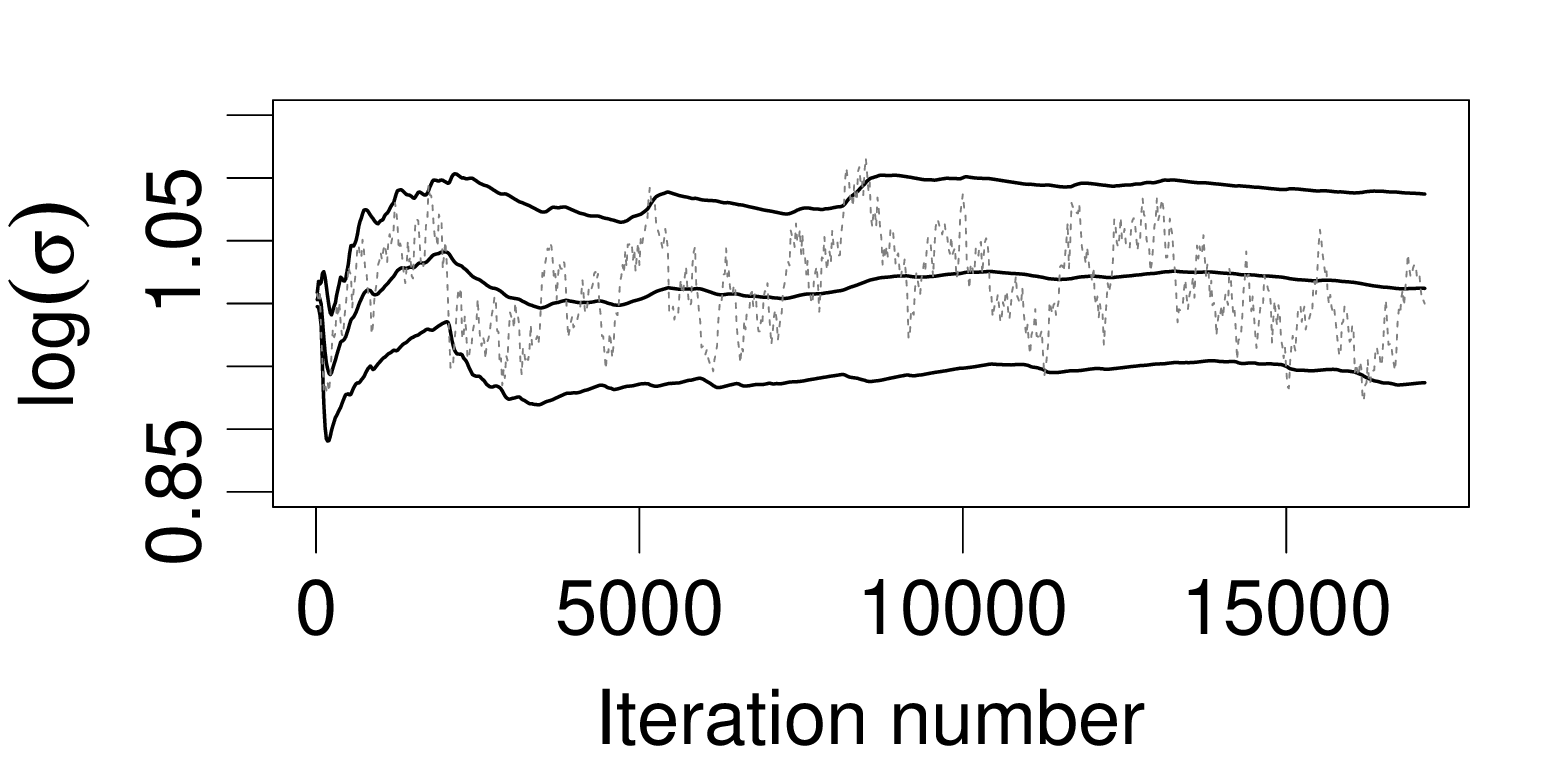} \includegraphics[width=0.49\columnwidth]{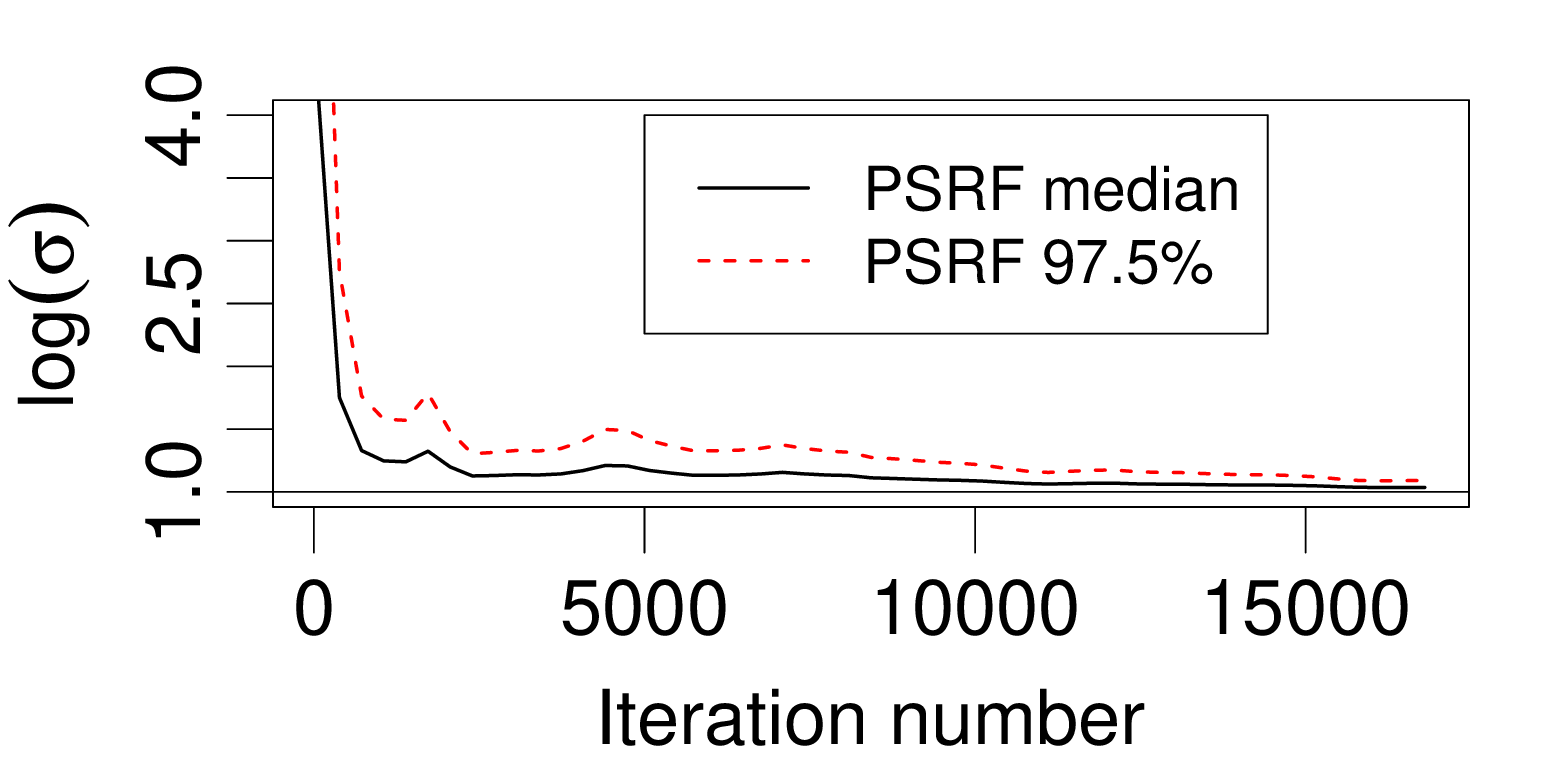}}
\centerline{\includegraphics[width=0.49\columnwidth]{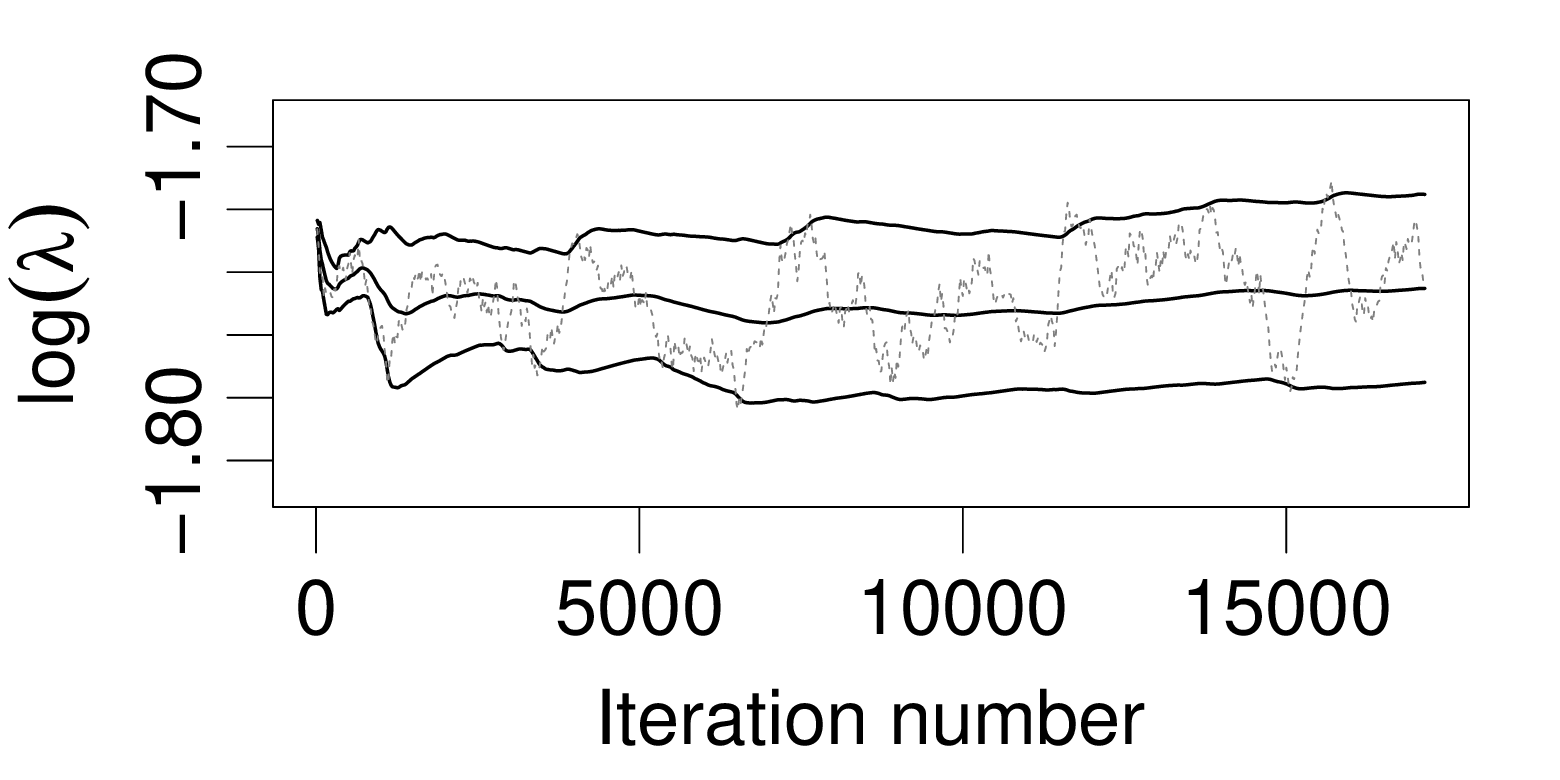} \includegraphics[width=0.49\columnwidth]{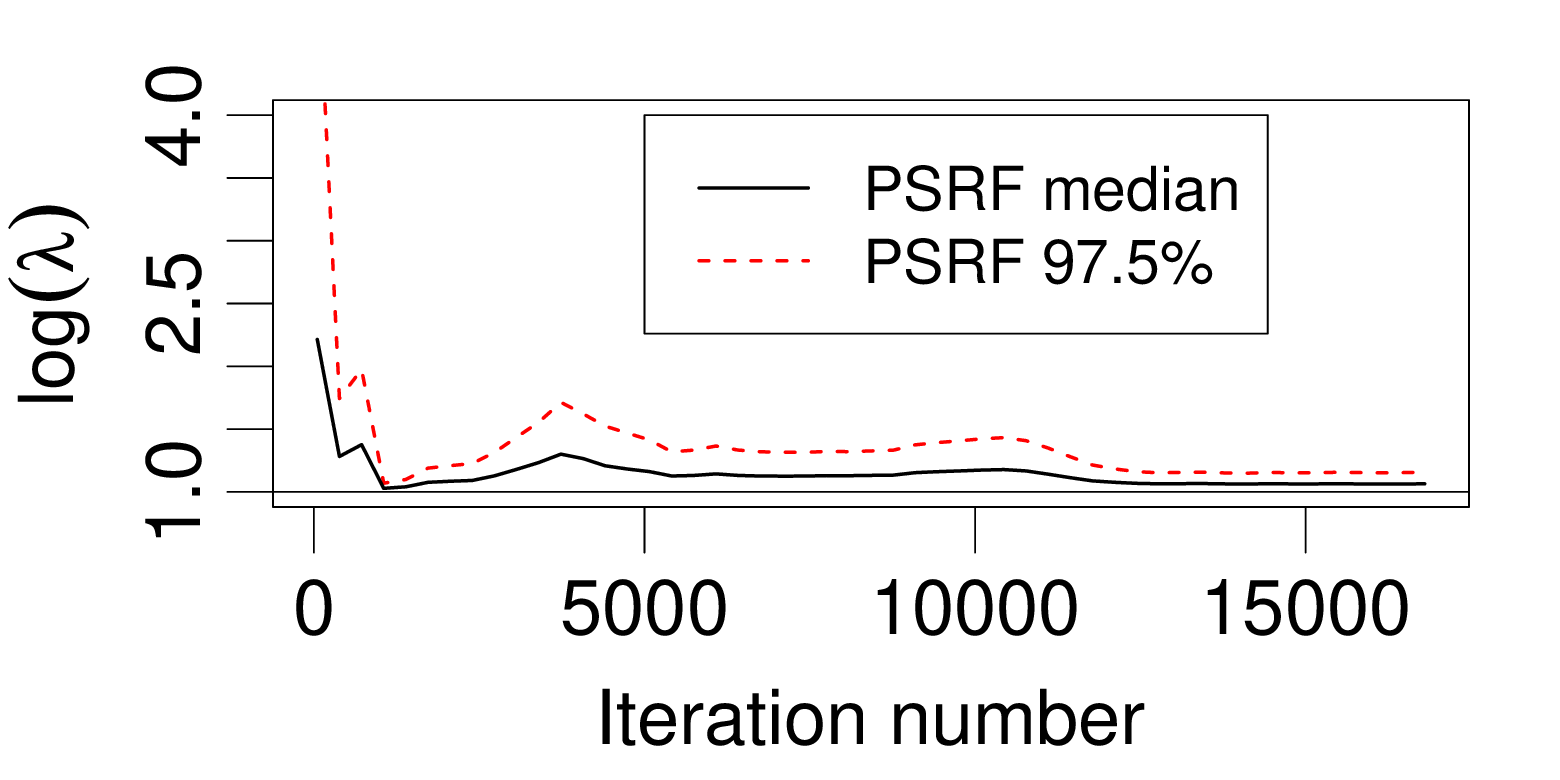}}
\centerline{\includegraphics[width=0.49\columnwidth]{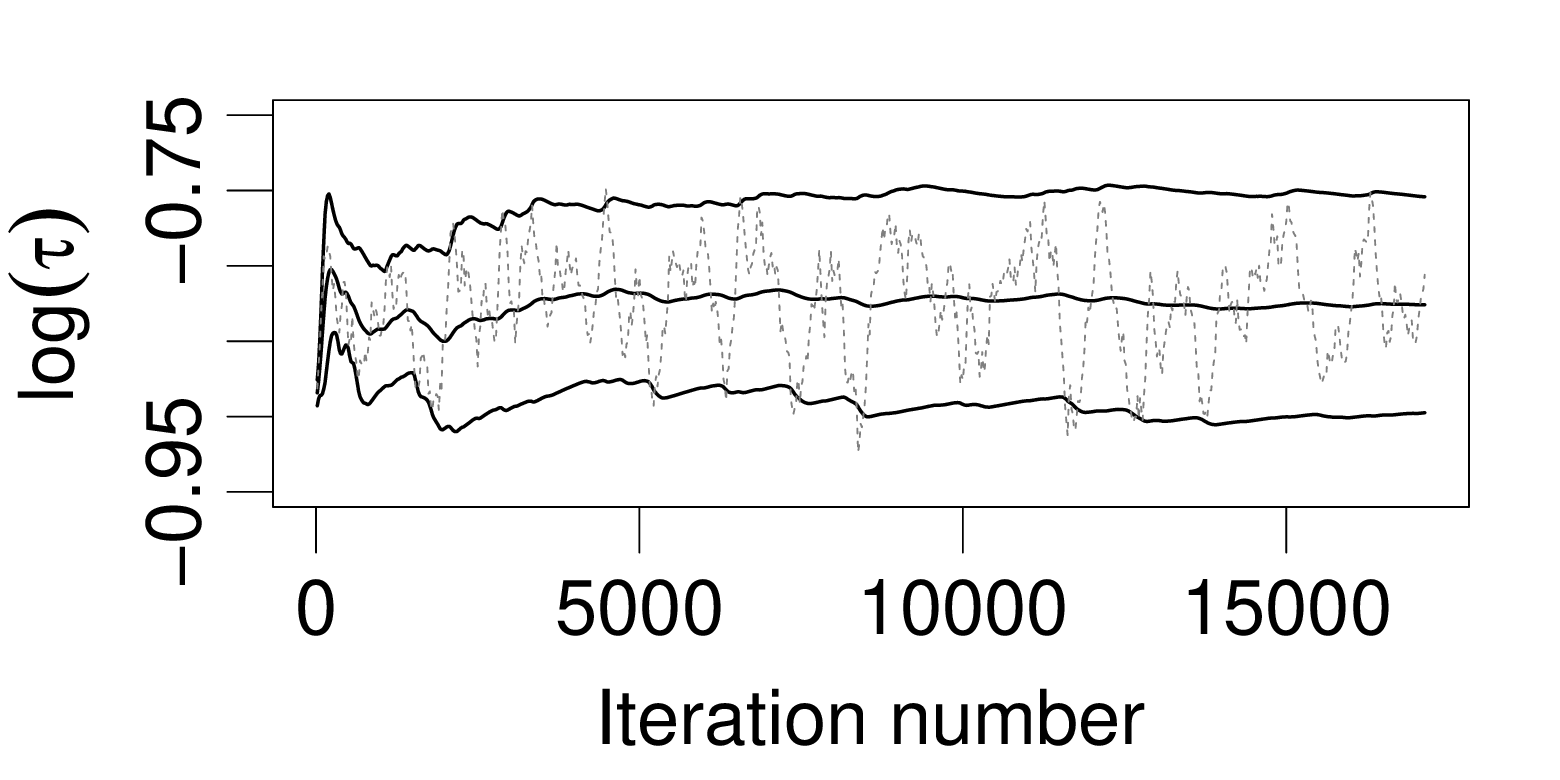} \includegraphics[width=0.49\columnwidth]{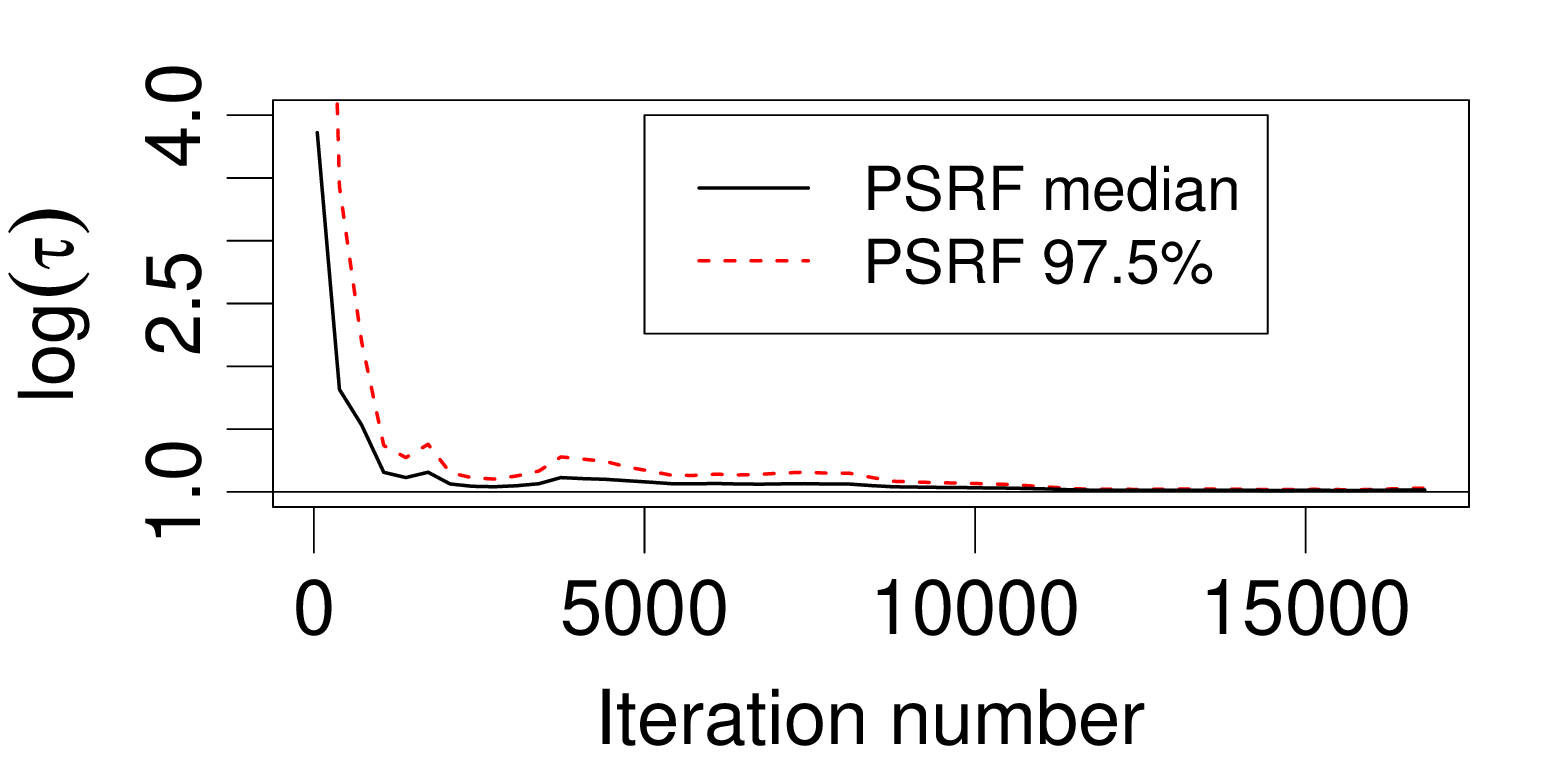}}
\caption{Census data ($n=22,784$) - Left panel: Running mean and plus/minus two standard deviations (black) and trace of one chain of SGLD (gray). Right panel: Convergence analysis of SGLD with ULISSE reporting the PSRF computed over five chains.}
\label{fig:final:census}
\end{center}
\vskip -0.2in
\end{figure}

We now present the application of SGLD with ULISSE to a data set where it is not possible to run any MCMC algorithm with exact computation of the marginal likelihood on a conventional desktop machine.
This data set contains data collected as part of the 1990 US census.
In this study, we used the 8L data set\footnote{\url{www.cs.toronto.edu/~delve/data/}} where the regression task associates the median house price in a given region with demographic composition and housing market features ($n=22,784$ and $d=8$). 
We kept the same experimental conditions as in the case of the Concrete data, except that $\varepsilon_t$ was chosen to decrease from $5 \cdot 10^{-2}$ to $5 \cdot 10^{-6}$ to cope with the larger gradients obtained for this data set, and the preconditioner $M$ was estimated based on the MAP on one-thousand data points.
%% Fig.~\ref{fig:final:census} reports the running mean and the running mean plus/minus two standard deviation for the three covariance parameters, along with the traces.
The running statistics for the three parameters for one chain are reported in Fig.~\ref{fig:final:census}, along with the PSRF computed across five chains, which shows that convergence was reached after few thousand iterations.

SGLD with ULISSE was run on a desktop machine with an eight core (i7-2600 CPU at 3.40GHz) processor, and an NVIDIA GeForce GTX 590 graphics card (released in 2011).
The two GPUs in the graphics card are used to carry out CMVPs.
With this arrangement, we were able to draw roughly ten thousand samples per day from the posterior distribution over covariance parameters.
SGLD yields an effective sample size of roughly $0.1\%$, and it can draw one independent sample every $2.4$~hours.

\section{Conclusions} \label{sec:conclusions}

This paper presented a novel way to accurately infer covariance parameters in GPs.
The novelty stems from the combination of stochastic gradient-based inference and a fast unbiased solver of linear systems.
The results demonstrate that it is possible to carry out inference of GP covariance parameters over a data set comprising about $23$ thousand input vectors in a day on a desktop machine with standard hardware.
The proposed methodology can exploit parallelism in computing covariance matrix-vector products, so there is an opportunity to scale ``exact'' inference (in a Monte Carlo sense) to even larger data sets.
We are not aware of any method that is capable of carrying out full quantification of uncertainty of GP covariance parameters on such large data sets without imposing special structures on the covariance or reducing the number of input vectors.
These results are important not only in Machine Learning, but also in areas where quantification of uncertainty is of primary interest and GPs are routinely employed, such as calibration of computer models \cite{Kennedy01} and optimization \cite{Jones98}.

The results reported in this paper, although promising, indicate some directions for improvements.
SGLD requires the tuning of a preconditioning matrix $M$.
Choosing $M$ to be similar to the covariance of the posterior speeds up convergence of SGLD when it reaches the Langevin dynamics phase.
However, $M$ also affects the scaling of the gradient in the proposal.
During the first phase of SGLD this might not be optimal, and %, and might not be optimal.
ideally, gradients should be scaled in a way similar to AdaGrad \cite{Duchi11}.
In \cite{Welling11}, it was possible to establish a connection between the covariance of the gradients, the Fisher Information, and $M$ due to the fact that stochastic gradients are computed on subsets of the data.
We were unable to do so for GPs due to the different way stochasticity is introduced in the computation of the gradients. %link the variance of the gradient with $M$.
Despite this complication, we demonstrated that it is still possible to obtain convergence to the posterior distribution over covariance parameters in a reasonable number of iterations, which is of ultimate importance in any inference task.

We are currently investigating the application of SGLD to automatic relevance determination covariances and the possibility to extend our proposal to scale inference for other GP models, e.g., GP classification and GPs for spatio-temporal data.
%% In these cases, the conditioning of the covariance matrix would be much poorer than GP classification and preconditioning techniques would need to be employed.
Other interesting aspects to explore would be the introduction of mixed precision calculations within the CG algorithm to improve convergence and computation speed as presented, e.g., in \cite{Jang11,Cevahir09,Baboulin09}.

\newpage

\section*{Acknowledgments} 

MF gratefully acknowledges support from EPSRC grant EP/L020319/1.

%% \textbf{Do not} include acknowledgements in the initial version of
%% the paper submitted for blind review.

%% If a paper is accepted, the final camera-ready version can (and
%% probably should) include acknowledgements. In this case, please
%% place such acknowledgements in an unnumbered section at the
%% end of the paper. Typically, this will include thanks to reviewers
%% who gave useful comments, to colleagues who contributed to the ideas, 
%% and to funding agencies and corporate sponsors that provided financial 
%% support.  

%% \nocite{langley00}

%% \bibliographystyle{my_icml2015}
%% \bibliography{filippone}

\end{document}